\documentclass[journal=jacsat,manuscript=article]{achemso}

\usepackage[version=3]{mhchem}
\usepackage{physics}
\usepackage{subcaption}
\usepackage{hyperref}
\usepackage[dvipsnames,table,xcdraw]{xcolor}
\usepackage[utf8]{inputenc}
\usepackage{amsmath}
\usepackage{bm}

\author{Mikołaj Martyka}
\affiliation[UW]
{University of Warsaw, Faculty of Chemistry, Krakowskie Przedmieście 26/28 00-927, Warsaw, Poland}
\author{Joanna Jankowska}
\email{joanna.jankowska@uw.edu.pl}
\affiliation[UW]
{University of Warsaw, Faculty of Chemistry, Krakowskie Przedmieście 26/28 00-927, Warsaw, Poland}
\author{Pavlo O. Dral}
\email{dral@xmu.edu.cn}
\affiliation[XMU]{State Key Laboratory of Physical Chemistry of Solid Surfaces, College of Chemistry and Chemical Engineering, and Fujian Provincial Key Laboratory of Theoretical and Computational Chemistry, Xiamen University, Xiamen, Fujian 361005, China}
\alsoaffiliation[UMK]
{Institute of Physics, Faculty of Physics, Astronomy, and Informatics, Nicolaus Copernicus University in Toru\'n, ul. Grudziadzka 5, 87-100 Toru\'n, Poland}
\alsoaffiliation[UMK2]
{Institute of Advanced Studies, Nicolaus Copernicus University in Toruń, ul. Wileńska 4, 87-100 Toruń, Poland}
\alsoaffiliation[Aitomistic]
{Aitomistic, Shenzhen 518000, China}

\title
  {OMNI-P3x universal neural network potential for large-scale computational photochemistry workflows}

\AtBeginDocument{\singlespacing}

\begin{document}

\begin{abstract}
Computational photochemistry supports the design and understanding of new materials. However, the high cost of excited-state quantum-mechanical (QM) calculations makes screening large molecular libraries prohibitively expensive. To address this limitation, we introduce OMNI-P3x, a universal machine-learned interatomic potential (MLIP) for singlet ground- and excited-state simulations of neutral and charged organic molecules. Trained on both ground- and excited-state conformations, it performs complete computational photochemistry workflows with accuracy approaching TD-DFT. We demonstrate its practical applicability by screening a combinatorial library of 234,375 flavylium derivatives, where the pretrained model performs ground-state geometry optimization and predicts absorption properties without additional QM calculations. Following fine-tuning, we further screen 162,960 xanthene-based dyes, identifying new candidates for red, near-infrared and short-wavelength infrared emission. Furthermore, state-dependent neural charges, learned without reference atomic charges, provide an interpretable picture of intramolecular charge redistribution upon excitation.
\end{abstract}

\section{Introduction}
Computational photochemistry and photophysics accelerates materials design through \textit{in silico} prediction of absorption and emission spectra~\cite{Mori2024Vibronic, Palombo2022Neorhodopsin, Chibani2013BODIPY}, assessment of molecular photoreactivity and photostability~\cite{Adrion2023, Barbatti2010}, and analysis of light-induced reaction mechanisms~\cite{Curtolo2025, Feng2024}.  These capabilities facilitate the development of dyes~\cite{Sumita2022, Wang2020}, fluorescent probes for bioimaging~\cite{Gerstner2025}, molecular photoswitches~\cite{Hartinger2026,Hartinger2025}, organic light-emitting diodes~\cite{Chen2025,Aizawa2022,Omar2023}, solar cells~\cite{Gillett2021,Hart2026}, and molecular solar thermal energy-storage systems~\cite{Ganguly2025,Ganguly2026, HillersBendtsen2023}, where performance critically depends on their optical properties and photochemistry. By connecting molecular structure to photophysical and photochemical behavior, computational methods help rationalize experimental observations and guide the design of new candidate systems, narrowing the chemical space for costly experimental verification~\cite{Aizawa2022, Chen2025}.

Practically, however, such studies are hindered by the prohibitive computational cost of quantum-mechanical~(QM) methods for excited-state calculations.  Even with widely used, approximate approaches such as time-dependent density functional theory~(TD-DFT)~\cite{Runge1984,Casida1995}, the cost of evaluating excited-state properties grows rapidly with molecular size, and more computationally demanding wavefunction-based methods, such as ADC(2)~\cite{Schirmer1982ADC} or EOM-CC~\cite{Stanton1993EOMCC}, being even more expensive. Calculations that are feasible for individual chromophores become impractical to scale into hundreds of thousands of candidate molecules. Hence, theoretical studies are often limited to small sets or even single, representative molecules of a given class, and screening of vast libraries of molecular systems remains difficult~\cite{Hegely2022,Wu2026}.

Machine learning interatomic potentials~(MLIPs) provide a natural path for acceleration in such workloads, replacing QM calculations with machine learning~(ML) approximations. By pre-training on vast datasets of molecular properties, these models can, with or without domain-specific fine tuning, replace costly QM calculations with minimal error, at a fraction of the cost~\cite{Kovacs2025MACEOFF,anstine2025, smith2017ani1, Wood2025UMA}. For ground-state chemistry, usage of such surrogate models is becoming increasingly routine, allowing for both: simulating larger systems and a wider coverage of the studied chemical space~\cite{Yuan2026}. At the same time, however, such progress is much more limited when looking at ML-assisted photochemical simulations and workflows. Despite the development of many architectures for ML of excited-state properties, based mainly on neural networks~(NNs)~\cite{pyraimd, Spainn, SchNarc}, most applications have relied on models trained for specific, individual molecular systems~\cite{Westermayr2021,Westermayr2022, Li2021Photoisomerization, Hu2018KRR}.

Recent developments are pushing the frontier of excited-state ML methods into broad transferability across chemical space. Axelrod et al. developed a neural-network potential transferable in the domain of azobenzene photoswitches, allowing for their screening for properties such as switching quantum yields and thermal half-lives~\cite{Axelrod2022,Axelrod2023}. In a related study, Tiefenbacher et al. trained an MLIP for arylazopyrazole photoswitches and incorporated it into an automated workflow for locating minima, transition states, and singlet–triplet crossing points~\cite{Tiefenbacher2026}. This enabled screening of thermal relaxation rates and identification of structural features governing the competition between isomerization pathways. X-MACE~\cite{X-MACE,X-MACE-screening} enabled a similar screening of fluorescent-protein chromophore derivatives, introducing pretraining on a broader collection of chromophores before targeted fine-tuning for the systems of interest. The pretraining was shown to decrease the number of datapoints required to obtain an accurate model~\cite{X-MACE, X-MACE-screening}. Alongside, alternative approaches based on learning electronic Hamiltonians~\cite{Friederich2026,Juergens2026} and size-intensive architectures based on extremal pooling~\cite{Malosso2026} are being developed in the direction of transferability.

The MS-ANI (multi-state ANI) model architecture further showed that a single neural network can learn potential energy surfaces with sufficient accuracy to perform nonadiabatic molecular dynamics across distinct molecules~\cite{Martyka2025}. Expanding on this multi-state model, OMNI-P2x was the first universal neural network potential for excited-state property prediction~\cite{Martyka2026}, enabling out-of-the-box prediction of excitation energies and absorption spectra across a broad range of organic molecules. However, the model faced limitations stemming from its training data, with excited-state labels available only at ground-state equilibrium geometries, limiting its reliability for out-of-equilibrium structures required for tasks such as geometry optimizations or emission spectra prediction, crucial for photochemical workflows.

\begin{figure}[H]
\centering
  \includegraphics[width=\linewidth]{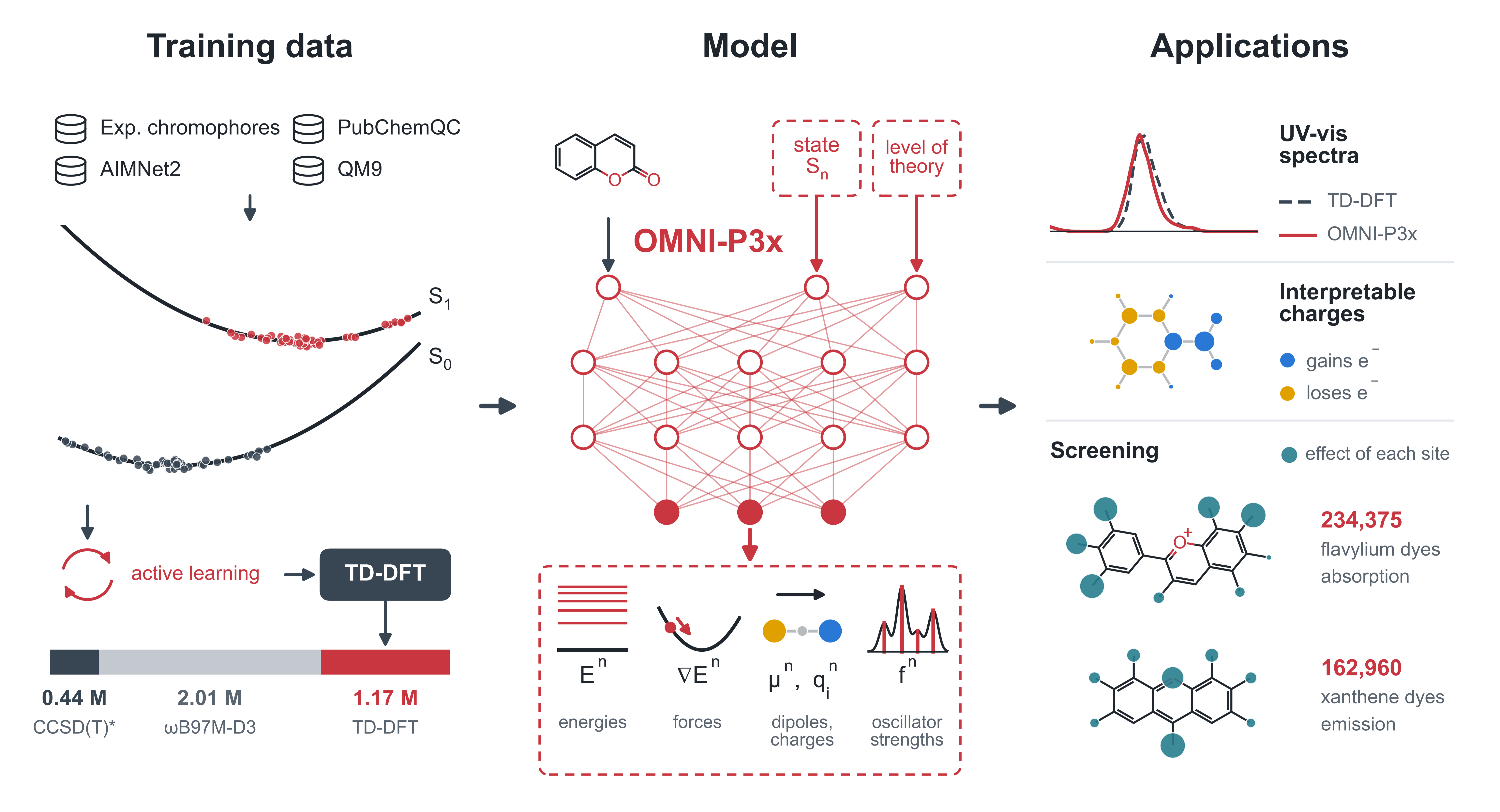}
  \caption{Overview of OMNI-P3x model development and application usage. Ground- and excited-state conformations of molecules from several databases were sampled, selected with active learning and labeled with TDA-$\omega$B97X-D/def2-SVP, giving a new dataset of 1.17 million structures. Combined with ground-state data at two further levels of theory, it was used to train OMNI-P3x, which predicts the excited-state properties needed for UV-vis spectra simulations, analysis of excited-state charges and high-throughput screening.}

  \label{fig:overview}
\end{figure}

Herein, we present OMNI-P3x, a universal MLIP for ground and excited-state simulations, built to enable large-scale photochemical workflows, with an overview of its design and capabilities provided in Figure~\ref{fig:overview}. It expands on its predecessor, OMNI-P2x, by providing broader elemental coverage and the ability to describe charged systems, as well as predicting state-dependent permanent dipole moments. It is trained on a newly-built training set, generated by sampling the ground and excited-state potential energy surfaces~(PESs), enabling geometry optimization in both, as well as subsequent prediction of absorption and emission properties. The model combines the established AIMNet2 architecture with multi-state~\cite{Martyka2025} and all-in-one~\cite{Chen2025_AIO}~(multifidelity) learning, incorporating additional ground-state reference data at multiple levels of theory, allowing it to retain accuracy when predicting ground-state properties. We assess its performance in ground-state and excited-state predictions, by comparing the results to both: the reference method, as well as theoretical best estimates throughout multiple test sets. We further show that the state-dependent neural charges, learned without reference atomic charges, provide an interpretable picture of charge redistribution between molecular units upon excitation. Such analysis could prove useful in designing materials which crucially depend on charge-transfer properties, such as  dye-sensitized solar cells~\cite{Alnoman2021} or donor--acceptor emitters for thermally activated delayed fluorescence~\cite{Uoyama2012}, but, so far, was only shown as a proof-of-concept capability of neural network potentials~\cite{Westermayr2020}.

Importantly, we show that OMNI-P3x can perform a full workflow required to predict photochemical and photophysical properties, from optimizing the ground or excited-state geometry through calculating frequencies and generating conformers for vibrationally resolved spectra, to predicting transition energies and intensities. Its practical applicability is shown by performing high throughput screening on a combinatorial library of 234,375 flavylium derivatives, where the pre-trained model is able to perform the entire task consisting of ground-state geometry optimization followed by absorption-spectrum prediction, with accuracy close to TD-DFT. We further apply fine-tuned OMNI-P3x to emission screening of a library of 162,960 xanthene-based candidate dyes, enabling us to propose novel emitter systems.

\section{Results and discussion}

\subsection{Model architecture and training}

The OMNI-P3x architecture is based on the AIMNet2 model and its pretrained weights (base model)~\cite{anstine2025}, extended to excited states \textit{via} the multi-state learning scheme developed in Ref~\citenum{Martyka2025}. It allows for efficient learning of an arbitrary number of electronic states and their properties in a shared neural network, with the electronic state encoded at the input layer to condition the entire network on the state of interest. Like its predecessor, OMNI-P3x utilizes an all-in-one~(multifidelity) training paradigm~\cite{Chen2025_AIO}, which enables it to learn efficiently across different QM levels of theory. This is achieved at the descriptor level of the NN, as shown in Figure~\ref{fig:training-data}a: two additional input features are appended to the atomic feature vector constructed for each atom: a scalar index $n$ of the electronic state $S_n$, e.g., $n=0$ for $S_0$ and $n=1$ for $S_1$, and a one-hot-encoded level of theory vector~$\mathbf{d}$. This operation is performed at each message-passing iteration used to refine the descriptor, forming the final AIM~(atoms-in-molecules) vector for a given electronic state. Depending on the choice of $\mathbf{d}$, the model predicts properties at the CCSD(T)$^*$/CBS and $\omega$B97M-D3(BJ)/def2-TZVPP levels for ground-state properties, or TDA-$\omega$B97X-D/def2-SVP level for ground- and excited-state properties. Throughout this work, we denote this choice as OMNI-P3x@CCSD(T)$^*$, OMNI-P3x@$\omega$B97M-D3(BJ), and OMNI-P3x@TDA-$\omega$B97X-D. AIMNet2 also performs a neural charge equilibration procedure, distributing atomic partial charges so that they sum to the total charge of the molecule $Q$, giving rise to a state-dependent, learned partial charge distribution.

OMNI-P3x features two output heads acting on the AIM~(atoms in molecules) vector, which are multi-layer perceptrons with 2 hidden layers of 128 neurons. The first one is an energy head, analogous to the base model, which obtains the final energy from summation of atomic contributions, $\varepsilon_i^n$, where $i$ is the index of a given atom, and a long-range electrostatic term, which is the Coulombic interaction of the neural partial charges. The second one is an oscillator strength head, which predicts the oscillator strength for an $S_0 \rightarrow S_n$ transition. Forces $\mathbf{F^n} =-\nabla E^n$ are obtained by automatic differentiation of the energy output. The partial charges are summed as $\sum r_iq_i^n$, where $r_i$ is the position of atom $i$, and $q_i^n$ is its neural charge of state $n$, to form the permanent dipole moment of electronic state $n$, $\bm{\mu}^n$. These four metrics are minimized jointly in the training process, using the loss function:
\begin{equation}
  \mathcal{L} =
    w_E \frac{\left\| E^{\mathrm{ML}} - E^{\mathrm{ref}} \right\|^2}{\sqrt{N_\mathrm{at}}}
  + w_F \left\| \mathbf{F}^{\mathrm{ML}} - \mathbf{F}^{\mathrm{ref}} \right\|^2
  + w_\mu \left\| \sum_i \mathbf{r}_i q_i^{\mathrm{ML}} - \boldsymbol{\mu}^{\mathrm{ref}} \right\|^2
  + w_f \left\| f^{\mathrm{ML}} - f^{\mathrm{ref}} \right\|^2 ,
  \label{eq:loss}
\end{equation}
where the superscripts ML and ref denote the model prediction and the reference value, $\|\cdot\|^2$ is the mean squared error, and $N_\mathrm{at}$ is the number of atoms. The loss weights for energies, forces, dipole moments and oscillator strengths were set to 1, 0.1, 0.1 and 0.5, respectively, selected by hyperparameter search and empirical evidence.

\subsection{Training set generation}
Having robust and high-quality training data is a prerequisite for constructing any successful MLP. However, due to the additional complexity of computational photochemistry~\cite{Dreuw2026}, the amount of available data on electronic excited states of molecules is still very limited. In particular, there is no available dataset that contains energy gradients for excited states, as well as properties for off-equilibrium structures across chemical space. The inclusion of the latter is particularly crucial, as performing tasks such as geometry optimizations or prediction of fluorescence spectra requires the MLP to be accurate far from the S$_0$ minimum. Hence, in this work, we curated a new excited-state dataset, by both sampling existing large-scale datasets, and generating data for new systems and their conformers.

\begin{figure}[H]
\centering
  \includegraphics[width=\linewidth,height=0.780\textheight,keepaspectratio]{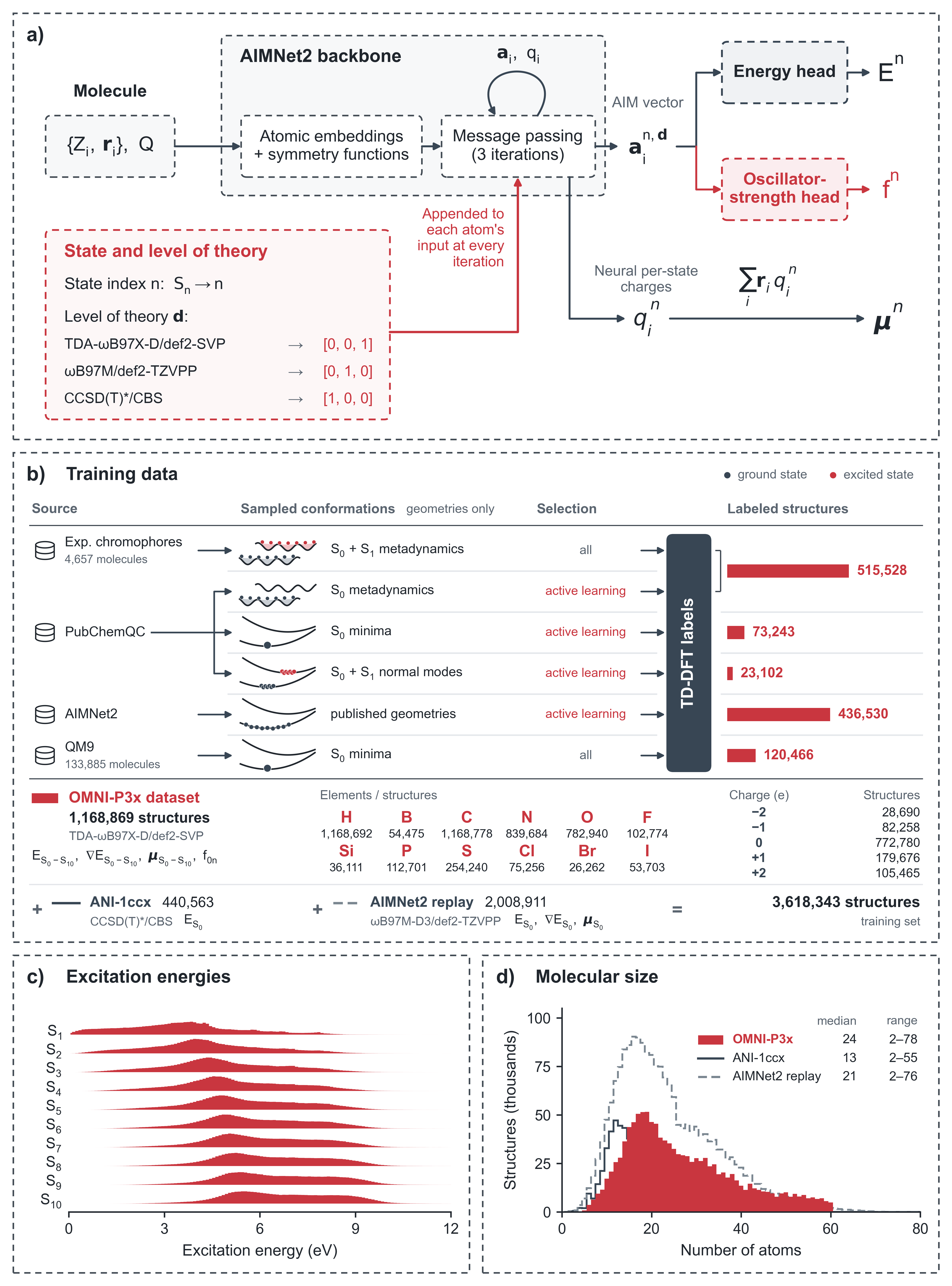}
  \caption{Model architecture and training data of OMNI-P3x.a) Neural network architecture of OMNI-P3x, based on the AIMNet2 backbone, extended to excited states. (b) Overview of the sampling procedure used to construct the new excited-state dataset, as well as its composition. (c) Distributions of excitation energies, $E(S_n)-E(S_0)$, for the 10 labeled excited states. (d) Distributions of the molecule size in the excited-state dataset and the auxiliary ANI-1ccx and AIMNet2 replay ground-state datasets. The reported mean and median refer to the excited-state dataset. All counts refer to molecular structures, including different conformations of the same molecule.}
  \label{fig:training-data}
\end{figure}

The starting point of the procedure was a database of SMILES strings of experimentally-known chromophoric systems of Joung et al~\cite{joung2020}. The molecules were converted to three-dimensional structures with RDKit,~\cite{rdkit} optimized with GFN2-xTB,~\cite{bannwarth2019gfn2} with conformational sampling performed using metadynamics in both the ground state and first excited state. The sampling procedure was applied to systems restricted to the set of 12 elements (H, B, C, N, O, F, Si, P, S, Cl, Br and I) with total charges between $-2$ and $+2$, as shown in Figure~\ref{fig:training-data}. Subsequent sampling stages were performed using uncertainty-based selection, which was quantified by disagreement between a main model, trained on energies and gradients, and an auxiliary model trained on energies only, as proposed in Refs.~\citenum{Martyka2025,hou2024}. Molecules were sampled from the PubChemQC~\cite{RN96}, AIMNet2~\cite{anstine2025}, and QM9 datasets~\cite{ramakrishnan2014}, with ground- and excited-state conformers generated using metadynamics and normal mode sampling. Only the molecular conformations were sampled in the procedure and, for all of them, reference labels were generated at the TDA-$\omega$B97X-D/def2-SVP level of theory, including energies of the ground state and ten lowest singlet excited states, together with transition dipole moments and oscillator strengths for the $S_0 \rightarrow S_n$ transitions. Energy gradients and permanent dipole moments were computed for $S_0$ and $S_1$, while their calculation for higher excited states was subject to a fixed computational budget per structure. An overview of the excited-state dataset sampling strategy is provided in Figure~\ref{fig:training-data}b, with full details in the Methods section. The newly-generated training set consisted of 1,168,869 TD-DFT datapoints with ground- and excited-state labels, with the distribution of their excitation energies shown in Figure~\ref{fig:training-data}c.

Like its predecessor, OMNI-P3x utilizes all-in-one learning~\cite{Chen2025_AIO} for simultaneous training across multiple levels of theory. Two additional datasets are used, the ANI-1ccx dataset~\cite{smith2020data}, with approximately 500,000 ground-state datapoints at the CCSD(T)/CBS fidelity, and the AIMNet2 dataset~\cite{anstine2025}, restricted to 10 \% randomly drawn conformations, for an additional 2,000,000 ground-state points labeled at $\omega$B97M-D3/def2-TZVPP, which acted as a replay set in the training procedure. The purpose of this extra ground-state data is mostly to enhance the performance of the model in the ground state, and retain the pre-trained AIMNet2 performance,  which is also crucial for photochemical tasks such as absorption spectra simulation. With the ground-state data included, the training set spans 3,618,343 conformations. The final composition of the dataset is shown in Figure~\ref{fig:training-data}b, with the distribution with respect to molecule size shown in Figure~\ref{fig:training-data}d, spanning systems between 2 and 78 atoms~(median 24).

\subsection{Performance of OMNI-P3x}
\begin{figure}[H]
\centering
  \includegraphics[width=\linewidth]{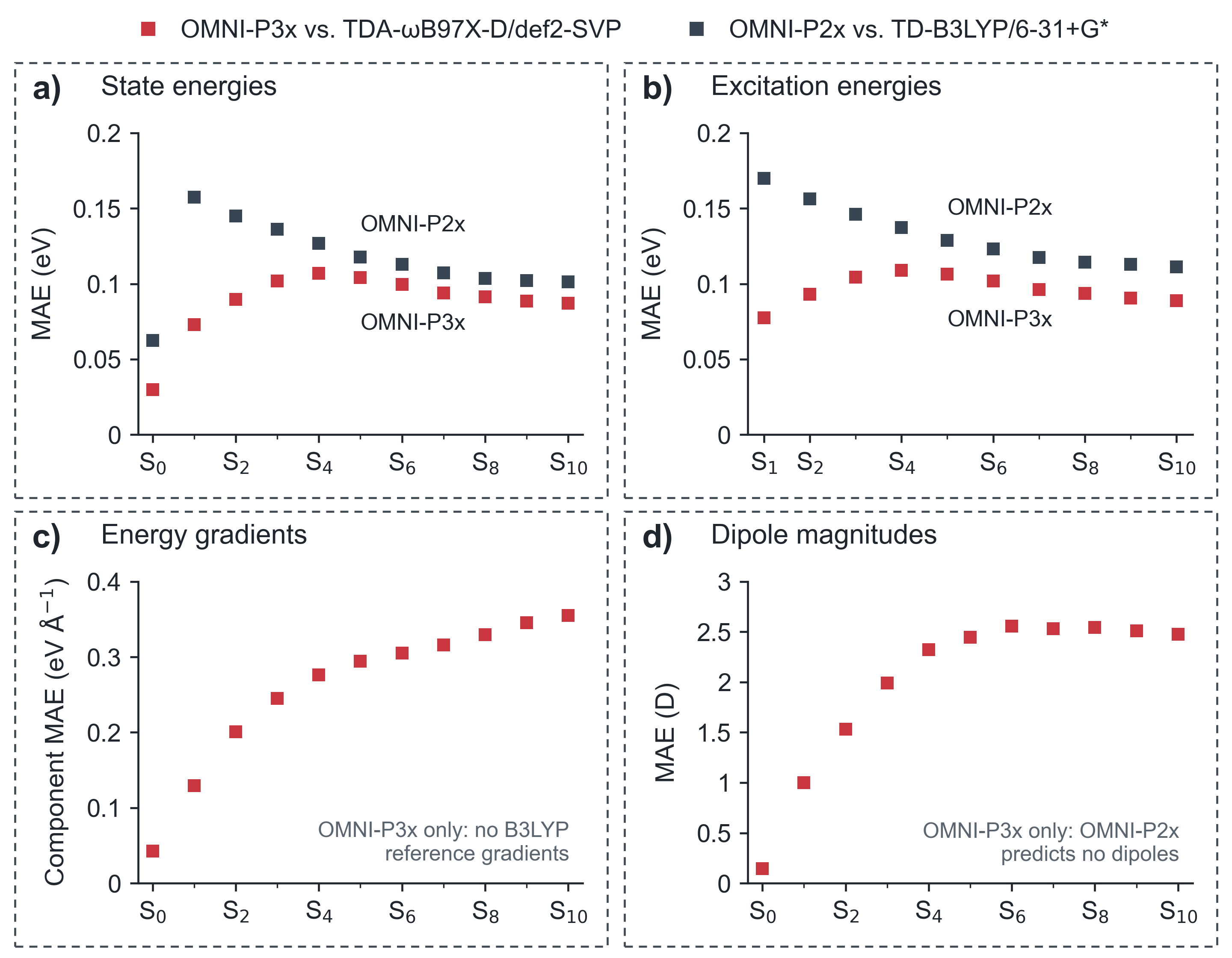}
  \caption{Per-state performance of OMNI-P3x and OMNI-P2x on PubChemQC data. Both models are compared to their own reference level of theory. a) Mean absolute error~(MAE) of electronic state energies, b) MAE of $S_0 \rightarrow S_n$ electronic transition energies, c) electronic state energy gradient MAEs, d) electronic state permanent dipole moment magnitude MAE. }
  \label{fig:pubchem-perstate}
\end{figure}

We first evaluate OMNI-P3x excited-state properties on a chemically diverse test set of 39,862 previously unseen molecules from PubChemQC (Figure~\ref{fig:pubchem-perstate}), allowing us to quantify the training and model extrapolation error. For the lowest excitation, $S_0\rightarrow S_1$, OMNI-P3x@TDA-$\omega$B97X-D achieves a mean absolute error (MAE) of 0.078~eV, which is less than half of the  0.170~eV of OMNI-P2x on the same set of molecules~(excluding the charged and B/Si/P/Br-containing systems unsupported by OMNI-P2x). Both models are compared to their own reference. Across the ten excited states the excitation-energy MAE rises from 0.078 eV for S1 to a maximum of 0.109 eV at $S_4$ and then decreases gradually to 0.089 eV at S10 (0.096 eV pooled over all states). Absolute state energies follow a similar trend: the MAE rises from 0.0298~eV for the ground state to 0.107~eV at $S_4$, then decreases to 0.087~eV at $S_{10}$.

Prediction of energy gradients and permanent dipole moments presents a greater challenge for the excited states. Their ground-state MAEs are 0.0424~eV/\AA{} per gradient component and 0.145~D, respectively, with larger errors for higher states. A particular difficulty is that states close in energy can have distinctly different electronic character and properties. For example, a charge-transfer state with a large dipole moment may lie almost energetically degenerate with a locally excited state having a much smaller dipole. An incorrect assignment of their ordering or character can therefore produce a large dipole error even when both state energies are predicted accurately. It is worth noting that this comparison error is not inherent to ML models, but can also be present when comparing adiabatically sorted properties between any two methods which, despite similar energies, can correspond to pairs of states with different electronic character.

\begin{figure}[H]
\centering
  \includegraphics[width=\linewidth]{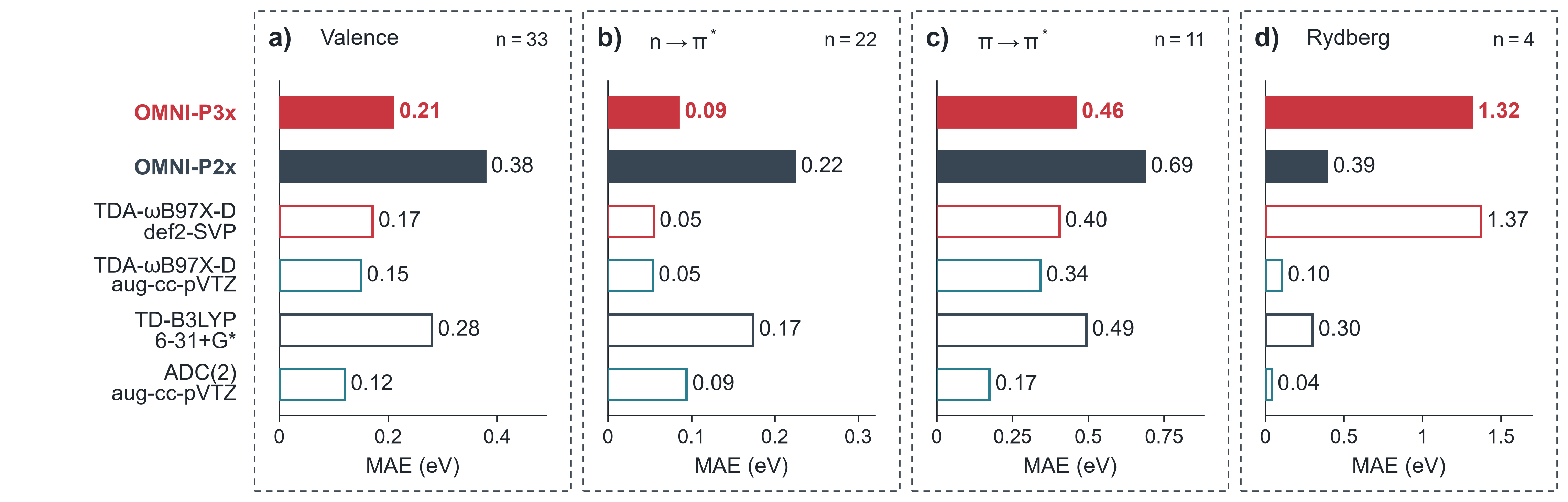}
  \caption{Prediction error of OMNI-P3x and other excited-state methods on the QUEST-3 and QUEST-5 datasets as MAE of $S_0 \rightarrow S_1$ transitions versus TBE. (a) All valence excitations; (b) $n \rightarrow \pi^*$ excitations; (c) $\pi \rightarrow \pi^*$ excitations; (d) Rydberg-type excitations.}
  \label{fig:vs_tbe}
\end{figure}

As the next step, we evaluate the accuracy of OMNI-P3x predictions versus a theoretical best estimate~(TBE), instead of the reference method, checking how accurate its predictions are compared to known methods for excited states. For this we use the QUEST-3~\cite{Loos2020QUEST3} and QUEST-5~\cite{Veril2021QUEST5} datasets, consisting of CC3~\cite{Koch1997CC3} optimized geometries of small and medium organic systems with excitation energies computed at a TBE level of theory. The nitroxyl molecule~(HNO) is excluded, as it is an inorganic system outside of the model's training domain. All methods are evaluated as single points at the published geometries, and the lowest predicted singlet excitation is compared with the TBE of the lowest singlet state listed for each molecule, for the 37 neutral molecules available to all methods, as OMNI-P2x does not have a charge input. A comparison between the TBE predicted $S_0 \rightarrow S_1$ excitation and OMNI-P3x@TDA-$\omega$B97X-D prediction is provided in Figure~\ref{fig:vs_tbe}. Overall, it achieves an MAE of 0.21 eV for all valence excitations, with considerably better performance for $n \rightarrow \pi^*$ excitations~(0.09 eV) than $\pi \rightarrow \pi^*$~(0.46 eV). It mirrors the performance of its training data very closely, with TDA-$\omega$B97X-D/def2-SVP achieving an overall MAE of 0.17 eV on valence excitations, with  significantly worse performance for $\pi \rightarrow \pi^*$ than $n \rightarrow \pi^*$. The error for valence excitations is not improved significantly when moving to a larger basis set, as  TDA-$\omega$B97X-D/aug-cc-pVTZ achieves an overall MAE of just 0.02 eV lower. The values can be compared to our previous generation model, OMNI-P2x, which has a valence MAE of 0.38~eV versus 0.28~eV of its training data, TD-B3LYP/6-31+G$^*$. Thus, the improvement in OMNI-P3x comes both from a more accurate level of theory and a better model that reproduces its reference more closely. OMNI-P3x struggles considerably more with Rydberg-type excitations, with an MAE of 1.32 eV, as shown in Figure~\ref{fig:vs_tbe}d. This is a property inherited from the small basis of the reference method~(MAE 1.37 eV), the error being almost completely removed in the aug-cc-pVTZ basis~(0.10 eV). OMNI-P2x, which was trained on TD-B3LYP/6-31+G$^*$, performs better due to the inclusion of diffuse functions in data generation, tracking the performance of its reference method. As a comparison, a higher-accuracy wavefunction-based method, ADC(2)/aug-cc-pVTZ scores a lower MAE of 0.12 eV on valence excitations, with the most notable improvement on $\pi \rightarrow \pi^*$ transitions with respect to DFT~(MAE 0.17 eV). It also achieves near-perfect accuracy on Rydberg excitations~(0.04 eV MAE).

Next, we evaluate how well OMNI-P3x performs in an end-to-end photochemical workflow, that is, optimizing a molecule in its ground or excited state and computing its electronic properties at the resulting minimum. This is of practical relevance, e.g., for predicting emission energies, which requires locating an excited state minimum, and for absorption-spectra predictions, which require computations at and around the ground-state minima. Such capability was not provided by any available universal MLIP, and the previous generation OMNI-P2x model's accuracy was restricted to single point calculations at ground-state only minima, that had to be located with a separate method.

\begin{figure}[H]
\centering
  \includegraphics[width=\linewidth,height=0.713\textheight,keepaspectratio]{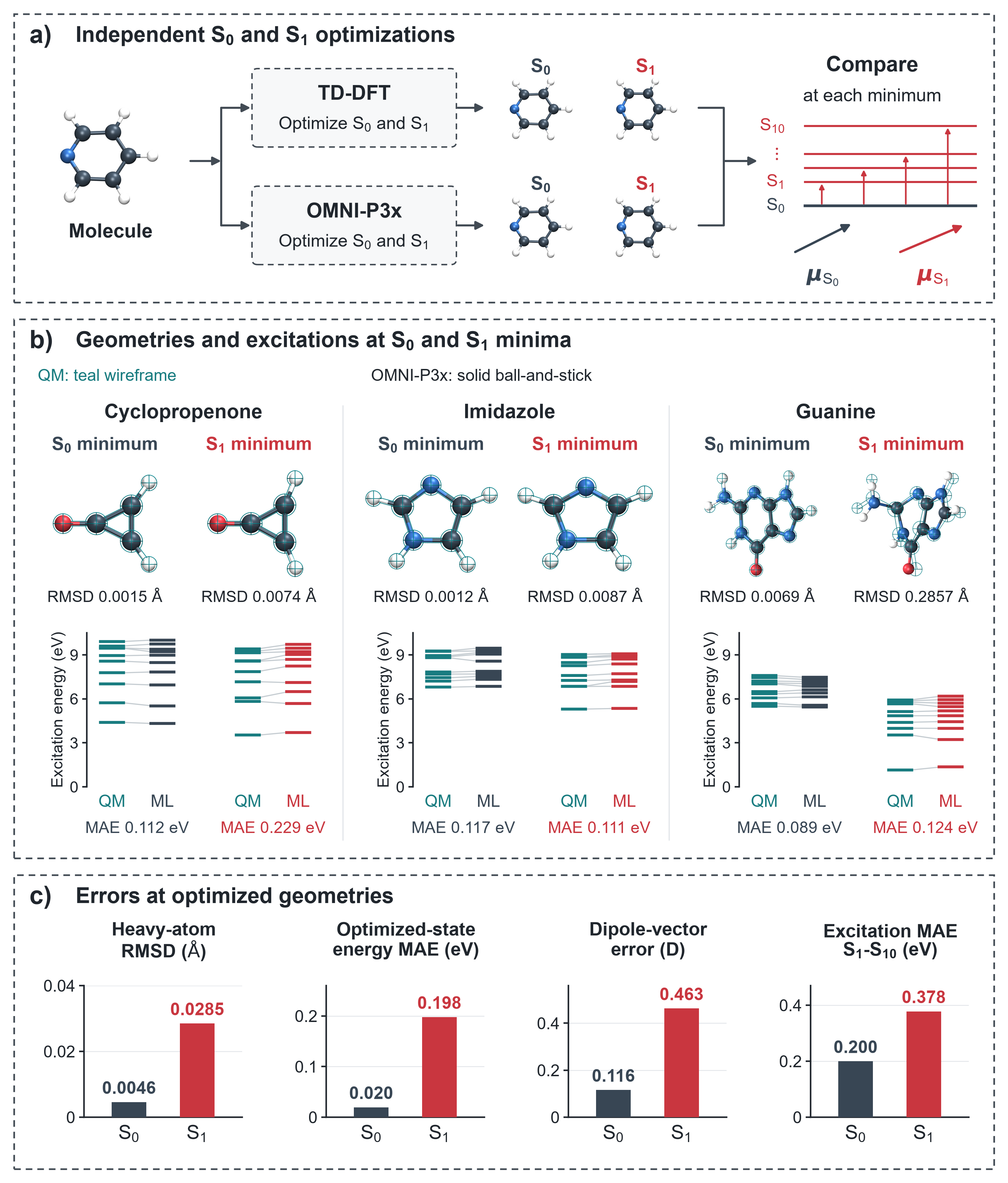}
  \caption{Performance of OMNI-P3x for geometry optimization and property prediction on the QUEST and Thiel benchmarks. a) Schematic depiction of the benchmark, consisting of $S_0$ and $S_1$ optimizations with TDA-$\omega$B97X-D and OMNI-P3x, and comparing properties at resulting minima. b) Example molecules from the benchmark, shown at their ML-optimized geometries~(solid ball-and-sticks), superimposed with QM results~(teal wireframe), along with their vertical absorption spectra, from OMNI-P3x (black at $S_0$ minimum, red at $S_1$ minimum), and TDA-DFT~(black). c) Mean heavy-atom RMSDs of the optimized minima, along with MAE of the total energy of the optimized state, optimized-state dipole-vector errors, and excitation-energy MAEs averaged over the first ten singlet excited states, for the benchmark molecules retained at each state. Black bars show statistics at $S_0$ minima, and red bars at $S_1$ minima.}
  \label{fig:benchmarks_quest_thiel_TDDFT}
\end{figure}

We select molecules serving as photochemistry benchmark systems, from the QUEST-3~\cite{Loos2020QUEST3}, QUEST-5~\cite{Veril2021QUEST5} and Thiel~\cite{Tuna2016} datasets~(62 molecules after pooling and deduplicating), between 2 and 14 heavy atoms. A given system is optimized with both OMNI-P3x@TDA-$\omega$B97X-D, and TDA-$\omega$B97X-D, in $S_0$ and $S_1$, and properties at the optimized minima are compared between methods. A schematic of the workflow is shown in Figure~\ref{fig:benchmarks_quest_thiel_TDDFT}. Again, nitroxyl~(HNO) is removed as an inorganic system out of the training domain, as well as pyrrole and thiophene from the $S_1$ evaluations, as their excited-state PES was dissociative at the TDA-$\omega$B97X-D/def2-SVP level of theory.

Examples of such workflows are shown in Figure~\ref{fig:benchmarks_quest_thiel_TDDFT}b, comparing ML-optimized geometries~(solid ball-and-stick) to reference QM optimization~(teal wireframes), for both $S_0$ and $S_1$, along with the corresponding vertical excitation energies. Error statistics for the test are shown in Figure~\ref{fig:benchmarks_quest_thiel_TDDFT}c. Overall, OMNI-P3x achieves a mean heavy-atom RMSD of 0.005 \AA~ for ground state optimizations, and 0.029~\AA~for the first excited state. The MAE in the final energy of the optimized state is 0.020~eV for $S_0$, and 0.198 eV for $S_1$. Error of the optimized state's dipole moment vector, taken as the length of the difference vector between the QM and ML values: $\lVert \Delta\vec{\mu} \rVert =
\lVert \vec{\mu}_{\mathrm{ML}} - \vec{\mu}_{\mathrm{QM}} \rVert$, after structural alignment, also rises from 0.116~D in the ground-state to 0.463~D following $S_1$ optimization. Finally, the MAE of the excitation energy, averaged over the entire $S_0 \rightarrow S_n$ spectrum, from $n=1$ to $n=10$ is equal to 0.200 eV and 0.378 eV, for $S_0$ and $S_1$ workflows, respectively.

The small errors of OMNI-P3x demonstrate that it reproduces the structure of the searched minima, as well as accurately predicts properties, such as energies and dipole moments at those minima. The larger errors following \(S_1\) optimization identify excited-state relaxation as the more demanding case, which is expected. Nevertheless, OMNI-P3x provides both the relaxed structures and their electronic properties within a single model, extending the practical scope of universal excited-state potentials from single-point predictions to complete ground- and excited-state optimization and property-evaluation workflows, examples of which we explore further in the manuscript.

\begin{figure}[H]
\centering
  \includegraphics[width=\linewidth,height=0.735\textheight,keepaspectratio]{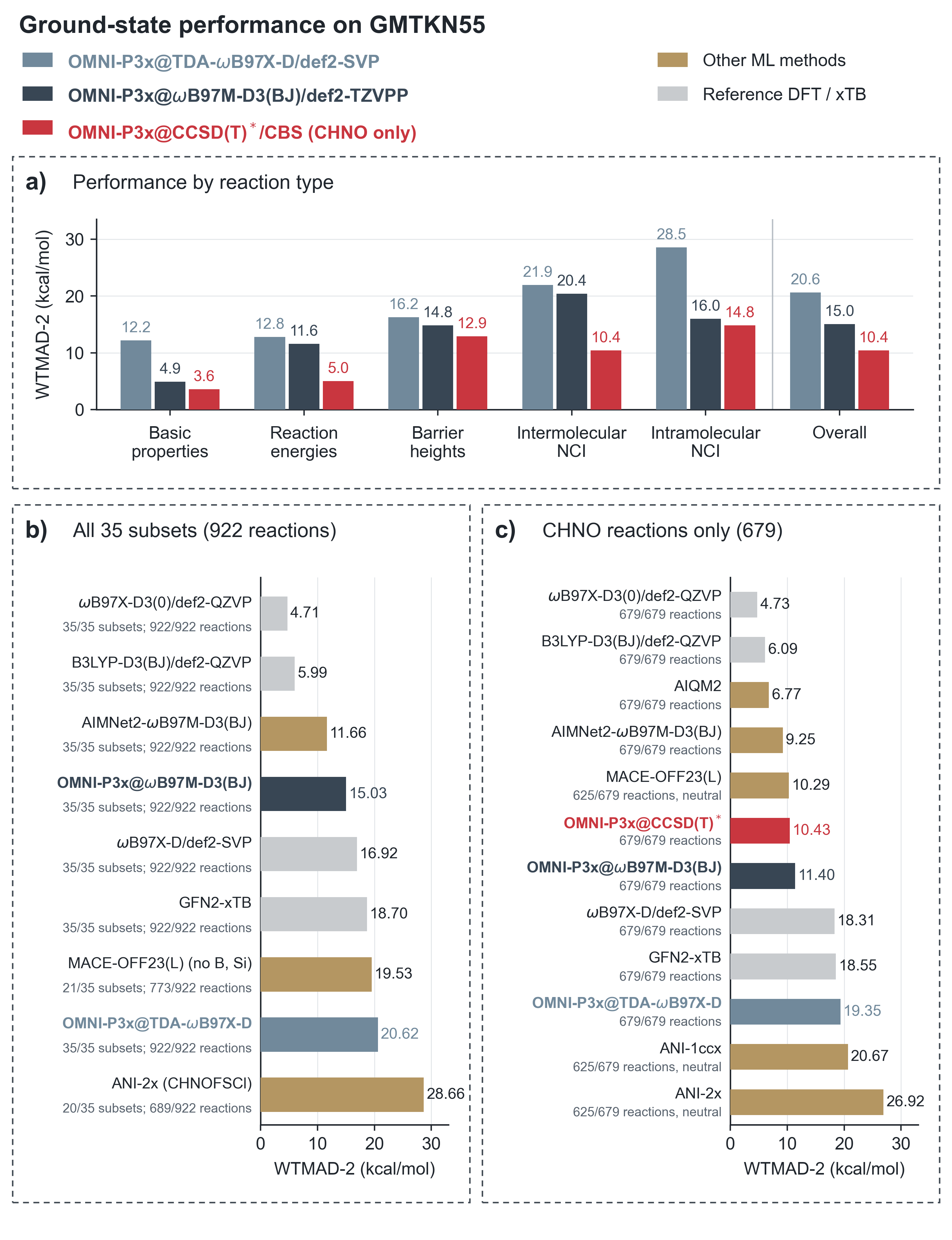}
  \caption{Ground-state performance of OMNI-P3x on the 35 GMTKN55 subsets fully within its closed-shell, 12-element training dataset. (a) WTMAD-2 by class for the three OMNI-P3x levels of theory, using 922 differences for each DFT level and 679 CHNO-only differences for CCSD(T). (b) Comparison using all supported reactions within this same 35-subset pool. The number of fully supported subsets and reactions is noted next to each method. The $\omega$B97X-D3(0)/def2-QZVP and B3LYP-D3(BJ)/def2-QZVP values are taken from the official GMTKN55 results~\cite{Goerigk2017}, all other results were calculated for the purpose of this work. ANI~\cite{smith2017ani1,Devereux2020ANI2x} and MACE were evaluated on neutral systems only. AIQM2 is included as a baseline.\cite{Chen2025AIQM2} (c) Comparison on the supported CHNO reactions, with coverage shown for each method; the pool contains 679 reactions.}
  \label{fig:WTMAD}
\end{figure}

Despite the model's main focus on the excited-state properties,  it retains a well balanced and usable accuracy in predictions of ground-state energetics as well. We assess this capability using GMTKN55, a widely used benchmark database comprising 55 subsets that span main-group thermochemistry, reaction barrier heights, and noncovalent interactions, serving as a standard tool to assess ground-state method accuracy~\cite{Goerigk2017}. The test set is narrowed down to 35 subsets consisting of closed-shell only examples, within the 12 fully supported elements of OMNI-P3x. Performance is quantified using the weighted total mean absolute deviation of type 2 (WTMAD-2) metric, as defined in Ref~\citenum{Goerigk2017}. Overall, OMNI-P3x@$\omega$B97M-D3(BJ) and OMNI-P3x@TDA-$\omega$B97X-D achieve WTMAD-2 values of 15.03 and 20.62 kcal$\cdot$mol$^{-1}$, respectively. OMNI-P3x@CCSD(T)$^*$ gives a WTMAD-2 of 10.43 kcal$\cdot$mol$^{-1}$ over the 679 supported CHNO reactions. The per-category error breakdown is shown in Figure~\ref{fig:WTMAD}a. The errors are lowest for basic properties, as well as reaction energies, while the model struggles considerably more with noncovalent interactions and reaction barrier heights.

A comparison of the ground-state performance of OMNI-P3x to other MLIPs and ground-state quantum chemistry methods is shown in Figure~\ref{fig:WTMAD}b, putting the above results in context. On their respective supported reaction sets, the WTMAD-2 of OMNI-P3x@$\omega$B97M-D3(BJ) lies between those of the AIMNet2 model~\cite{anstine2025} and MACE-OFF23(L)~\cite{Kovacs2025MACEOFF}, which achieve 11.66 and 19.53 kcal$\cdot$mol$^{-1}$, respectively. It also outperforms the semi-empirical GFN2-xTB method~(WTMAD-2 18.70 kcal/mol)~\cite{bannwarth2019gfn2}, and $\omega$B97X-D in the modest def2-SVP basis~(WTMAD-2 16.92 kcal/mol). Larger-basis DFT references remain significantly more accurate. Figure~\ref{fig:WTMAD}c shows the same comparison restricted to reactions involving CHNO elements only, which is the sub-set the   OMNI-P3x@CCSD(T)$^*$ supports. Figure~\ref{fig:WTMAD}c reports each method on its supported CHNO reactions.  The small-basis reference, $\omega$B97X-D/def2-SVP, has a WTMAD-2 of 16.92 kcal$\cdot$mol$^{-1}$ on the 35-subset pool and 18.31 kcal$\cdot$mol$^{-1}$ on the CHNO pool.

\subsection{UV-Vis absorption spectra}

\begin{figure}[H]
\centering
  \includegraphics[width=\linewidth,height=0.802\textheight,keepaspectratio]{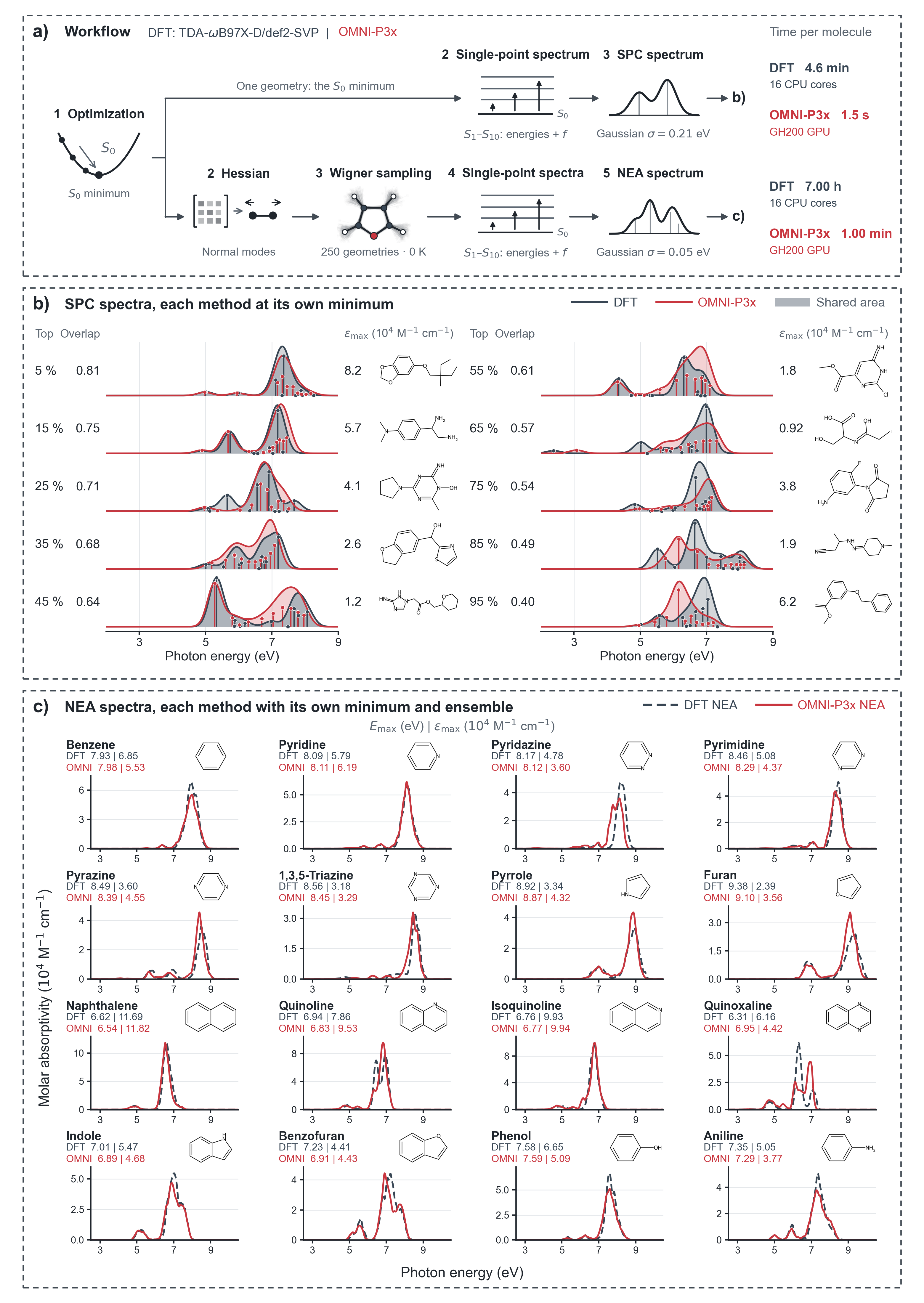}
  \caption{UV-Vis spectra simulations with OMNI-P3x. (a) Schematic depiction of the workflows for single-point convolution~(SPC) and nuclear ensemble approach~(NEA) spectra. along with the computational cost for OMNI-P3x and reference TD-DFT. (b) Simulated SPC spectra of test molecules from the PubChemQC dataset, shown as random examples drawn at different percetile rankings by the spectral overlap score.  (c) Simulated NEA spectra for 16 medium-sized organic molecules, with reference TD-DFT~(black lines) and OMNI-P3x~(red lines).}
  \label{fig:spectra-UVVIS}
\end{figure}

Next, we test how well OMNI-P3x reproduces UV-Vis simulations conducted with reference TD-DFT calculations. To do so, we select 1,000 random unseen structures from the PubChemQC test set, and perform ground-state geometry optimization and single-point spectrum prediction with TDA-$\omega$B97X-D/def2-SVP and OMNI-P3x@TDA-$\omega$B97X-D, as shown in Figure~\ref{fig:spectra-UVVIS}a (SPC branch). Spectra were broadened using a gaussian function with the broadening parameter set to 0.3 eV~(see Methods section for more details). Each method generates its own minimum, so the comparison assesses the full workflow. The spectra resulting from the ML and QM workflows are compared in Figure~\ref{fig:spectra-UVVIS}b, ranked by the spectral overlap (see Methods) between the reference and ML spectrum, from the top to the bottom 5\%. The mean overlap is 0.62, with the spectral overlap being a strict metric: a blue-shift of 0.2~eV alone gives a score of 0.58, while halving or doubling intensities gives 0.50. OMNI-P3x reproduces the transition energies very well (MAE 0.10~eV), and the optimized structures agree well with the reference (median heavy-atom RMSD 0.09~\AA). The overall shape of the spectra is usually well reproduced, with only 5\% of molecules below an overlap of 0.40, and the remaining discrepancies stem mainly from band intensities. A single simulation, including the optimization, takes on average 1.5~s with OMNI-P3x on a single GH200 GPU, compared with 4.6~min with the reference method on 16 CPU cores, giving a 180x speedup.

Nuclear ensemble approach~(NEA) is a method of simulating UV-Vis absorption spectra of molecules that allows for inclusion of vibrational effects in the spectrum, allowing for reproduction of second-order features such as peak splitting, or vibrationally allowed transitions~\cite{Bai2020,CrespoOtero2012}. The downside is that it requires extensive sampling of the ground-state potential energy surface, as well as performing single-point calculations of energies and oscillator strengths for hundreds or thousands of molecules. Machine learning allows for accelerating such simulations\cite{ML-NEA,MLNEA2, Westermayr2020, Dral2021}, and herein, we show that OMNI-P3x can be used for this purpose, without task-specific fine-tuning. An example NEA workflow is presented in Figure~\ref{fig:spectra-UVVIS}a, consisting of optimizing the ground state geometry of the system, computing its frequencies~(Hessian matrix), performing Wigner sampling to obtain an ensemble, performing single-point calculations for all ensemble members, and assembling the spectrum.

We test OMNI-P3x on 16 selected medium-sized, aromatic and heteroaromatic organic molecules, for which we compute the NEA spectra and compare with reference TD-DFT. Each method generates its own optimized minimum, normal modes and nuclear ensemble, so the comparison assesses the combined effects of structural relaxation, frequency calculation and excited-state property prediction. The spectra are broadened using gaussian functions with $\sigma$=0.05~(see Methods). Resulting spectra are compared in Figure~\ref{fig:spectra-UVVIS}c. OMNI-P3x reproduces the absorption maxima very well, with  a mean absolute difference between spectral maxima of 0.14 eV. The relative intensities of the peaks are also in most cases well reproduced, with the exception of quinoxaline, where OMNI-P3x wrongly predicts the higher energy peak to be more bright. Importantly, the simulations take on average 7~h per molecule with the reference method on 16 CPU cores, compared with 1~min with OMNI-P3x on a single GH200~GPU, giving a 420x speedup. These results show that OMNI-P3x can replace TD-DFT across this entire workflow, from the geometry to the spectrum, at a fraction of the cost.

\subsection{Interpretable excited-state charges}
All properties discussed so far derive from quantities that OMNI-P3x was directly trained on. The neural partial charges $q_i^n$ are different: the pre-trained model uses no reference atomic charges and learns the charges only indirectly, through the quantities they contribute to. These are the permanent dipole moment and the long-range Coulomb term of the energy, which both enter the loss function of Equation 1. Previous studies have shown that learned atomic quantities can provide insight into excited-state character even without an explicit wavefunction or electron density, as demonstrated with SchNarc~\cite{Westermayr2020} and OMNI-P2x~\cite{Martyka2026}. Here, we show that the neural charges follow
the change of the electron density upon excitation, so that their difference between the states, $\Delta q_i^n = q_i^n - q_i^0$, provides an interpretable picture of how the electrons move between molecular units. To this end, we compare the neural charges with the electron density itself, by integrating the relaxed TD-DFT difference density of each state over the atoms. As a benchmark we use the QUEST-6~\cite{Loos2021QUEST6} set of intramolecular charge-transfer (CT) excitations at its published geometries, total 26 electronic transitions in 16 distinct systems, excluding the HCL and two dipepetide systems. CT states are selected as a natural test case, involving a directed displacement of charge between well-defined molecular units. Since atomic charges are not uniquely defined, we mainly compare the charge summed over these units.

\begin{figure}[H]
\centering
  \includegraphics[width=\linewidth]{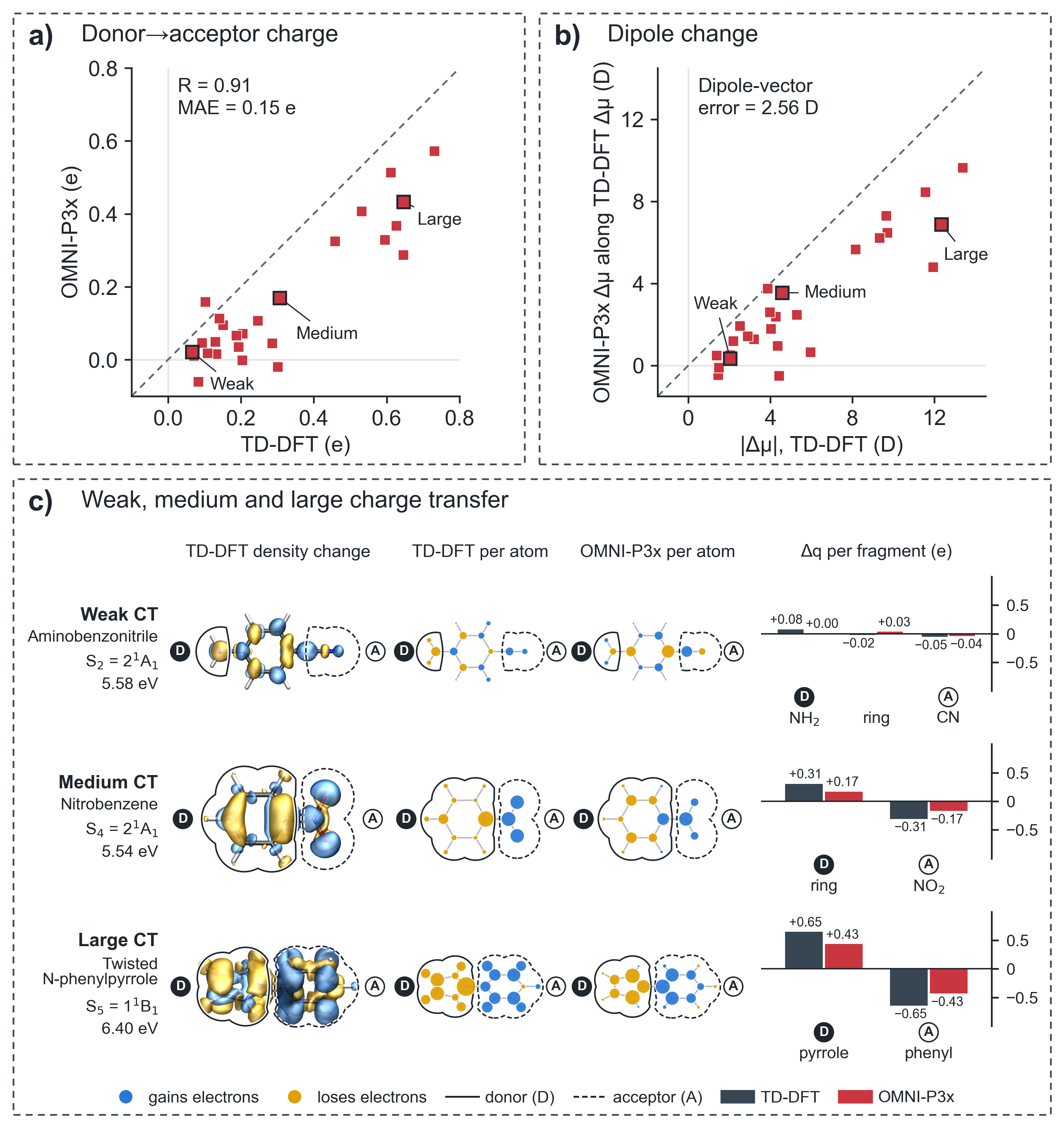}
  \caption{OMNI-P3x neural charges against TD-DFT for charge-transfer states of the QUEST 6 CT set. a) Charge quantity transferred from donor to acceptor, from the integrated relaxed difference density and from the OMNI-P3x atomic charges. b) Change of the dipole moment on excitation, with the OMNI-P3x vector projected on the TD-DFT one. c) Examples of excited states with weak, medium and large CT: relaxed difference density (plotted with isovalue of 0.002 a.u.), the same integrated per atom along with the change of the OMNI-P3x atomic charges, as well as the summed donor and acceptor charge changes.}
  \label{fig:quest-charges}
\end{figure}

As shown in Figure~\ref{fig:quest-charges}a, the neural charges follow the TD-DFT density change, with a correlation coefficient R=0.91 between OMNI-P3x and TD-DFT in the quantity of the charge transferred from the donor to the acceptor, and its direction is correct for 23 of the 26 states.  Better agreement is obtained for strong CT states, than systems with weak CT character. The change of the dipole moment between the ground-state and CT state, shown in Figure~\ref{fig:quest-charges}b has a MAE of 2.56~D, and good correlation with the TD-DFT predictions (R=0.91). Three typical cases~(weak, medium and large) CT are illustrated in Figure~\ref{fig:quest-charges}c.  These are the transfer from the amino to the cyano group in aminobenzonitrile, from the ring to the nitro group in nitrobenzene, and from the pyrrole to the phenyl ring in twisted \textit{N}-phenylpyrrole. In all cases, OMNI-P3x correctly identifies the  donor and acceptor, and predicts the qualitative character of the CT state, while systematically underestimating the quantity of the transferred charge. Thus, while the neural charges of the pre-trained model do not fully capture the quantity of the transferred charge, they describe the qualitative character of these CT states very well, despite being learned without any charge reference.

\subsection{High-throughput screening of flavylium dyes}

\begin{figure}[H]
\centering
  \includegraphics[width=\linewidth,height=0.713\textheight,keepaspectratio]{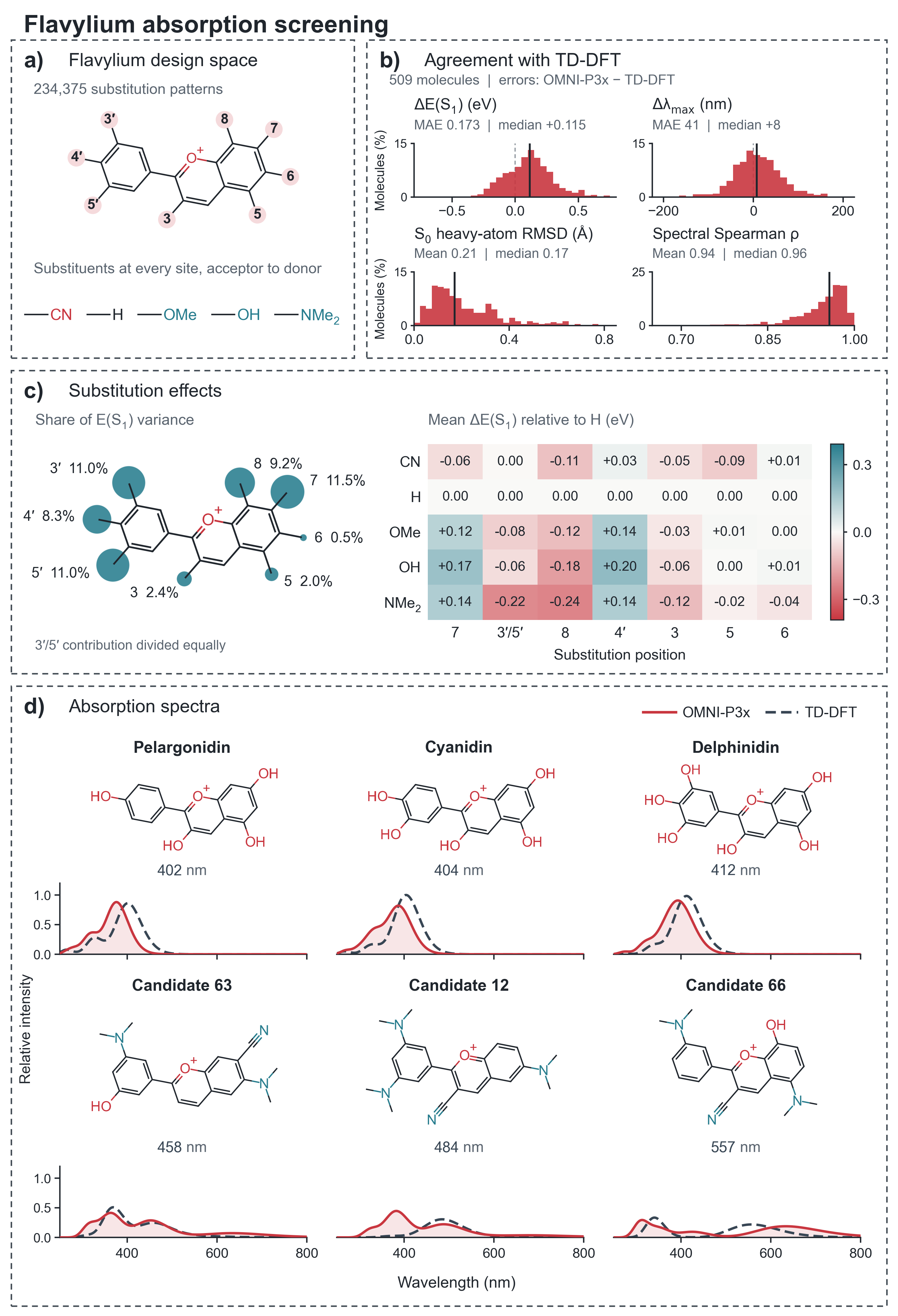}
  \caption{Screening of the flavylium design space with OMNI-P3x. Panel a shows the molecular scaffold, along with the 8 possible substitution sites, as well as the 5 screened substituents. Agreement of OMNI-P3x and reference TD-DFT screening outcomes is shown in panel b, described by the error in the $S_0/S_1$ energy gaps, the wavelength of the first absorption maximum, heavy-atom RMSD and the Spearman rank correlation. Substituent effects on the $S_0/S_1$ energy gap are shown in panel c, as the share of variance~(see Methods) explained by each site. Circle areas are proportional to each site's share of the variance. Effects of each substituent at each position, quantified as the mean change in the $S_0/S_1$ gap relative to an -H substituted system, are shown on the right. Panel d shows simulated spectra, with OMNI-P3x~(solid red line) and TD-DFT~(dashed black line) for three natural product anthocyanidins (top), and three new candidate systems.}
  \label{fig:absorption-screening}
\end{figure}

High-throughput screening is another particularly promising application of universal MLPs, as they can quickly process vast libraries of molecular structures. Here, we demonstrate that OMNI-P3x can be used for this task out of the box, in an example screening of the absorption properties of organic dyes. For this purpose, we have selected the flavylium cation, which is the core of numerous natural product dyes, such as anthocyanins, with related dyes finding applications in photochromic materials and fluorescent probes for bioimaging~\cite{Cruz2022Flavylium,Sun2022Flavylium}. It is depicted in Figure~\ref{fig:absorption-screening}a, along with the studied substitution pattern, spanning 5 choices of varying donor-acceptor character. Using OMNI-P3x, we were able to study the full design space spanning 234,375 systems. Importantly, the ML model was used for the entire workflow: both the ground-state minimum optimization, as well as the UV-Vis spectrum prediction at that minimum. The MLP provided a speed-up of a factor of over 500, with the properties of a benchmark dye taking on average 4.7 s on a single GH200 GPU to compute, versus almost 45 minutes for TD-DFT calculations, on 16 CPU cores. In total, the screening of the full design space took 435 GPU-hours, and would require about 2.7 million CPUh to perform with DFT.

The performance of OMNI-P3x is evaluated in Figure~\ref{fig:absorption-screening}b, on a set of 509 test molecules, composed of 500 randomly selected systems from the design space, as well as 9 known reference compounds based on the flavylium cation. Encouragingly, OMNI-P3x achieves an MAE of 0.17~eV in terms of the $S_0/S_1$ energy gap, and an MAE of 41 nm~(median signed error 8~nm) in predicting the lowest absorption maximum of the dyes. The optimized structures agree well with the reference, having a median heavy-atom RMSD of only 0.170~\AA. Excellent agreement is noted for the spectra as well, with a median Spearman correlation coefficient (SCC) of 0.96 between reference TD-DFT and OMNI-P3x simulations for a given system. Importantly, the metrics represent discrepancies between the full workflows, and the end results of optimization and spectrum simulation are compared between the two methods.

Analysis of a full combinatorial library of the flavylium cation allows us to obtain direct insights into the effects of each substituent at each site, as well as the relative importance of each substitution site, taken as the share of the $S_0/S_1$ transition energy variance explained by this site. In particular, positions 8, $3'$/$5'$~(analyzed together due to the symmetry of the system), 7 and 4' stand out as the most important,  with the remaining three having a smaller effect on the final $S_0/S_1$ energy gap of the system. In terms of specific substituents, electron donors at positions 8 and $3'$/$5'$, which are conjugated with the positive charge on the oxygen, provide a significant red-shift, with a particularly strong effect in the case of the dimethylamino group, while donors at the $4'$ and 7 sites introduce a blue-shift. The electron-withdrawing cyano functional group on average introduces a decrease in the absorption energy of the system. For the $3'$/$5'$ and 8 positions, dimethylamino substitution introduces a strong red-shift, which is reversed in the 7 and $4'$ positions.

Figure~\ref{fig:absorption-screening}d shows the simulated single-point convolution absorption spectra of the studied dyes. In the top row, OMNI-P3x predictions are compared to TD-DFT results for three known natural products: pelargonidin, cyanidin and delphinidin, showing an excellent agreement between the MLP and reference. Furthermore, our screening was able to identify novel flavylium derivatives with a significantly more red-shifted absorption. For the 100 candidates with the lowest $S_0 \rightarrow S_1$ excitation energies and at most 4 substituents, as predicted by OMNI-P3x, reference TD-DFT/$\omega$B97X-D calculations were carried out. Three example structures, along with their spectra, are shown in the bottom row of Figure~\ref{fig:absorption-screening}d.

\subsection{High-throughput screening of xanthene-based emitters}

\begin{figure}[H]
\centering
  \includegraphics[width=\linewidth,height=0.690\textheight,keepaspectratio]{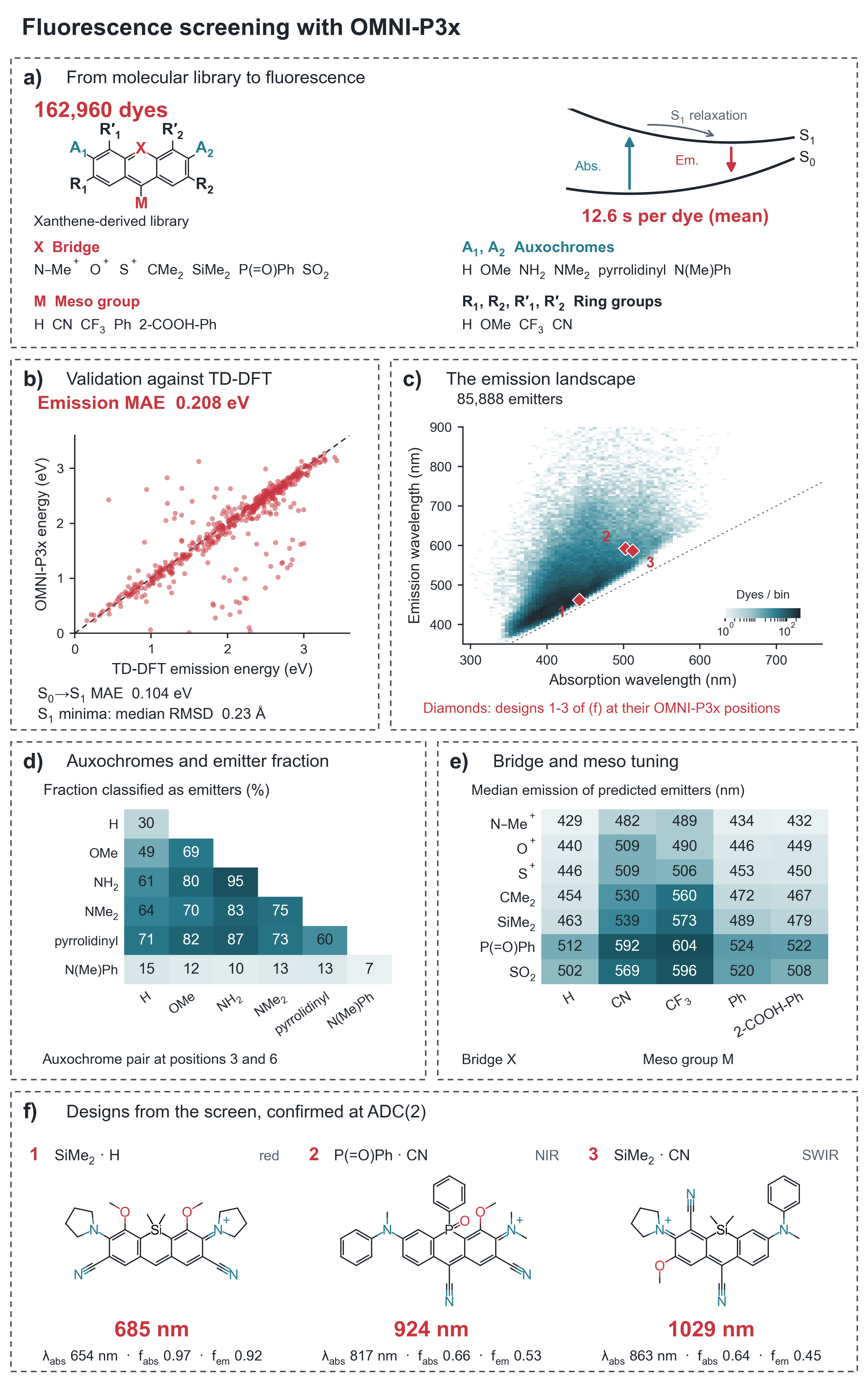}
  \caption{Fluorescence screening of a xanthene-based molecular library with fine-tuned OMNI-P3x. Wavelengths in panels c--e are OMNI-P3x@TDA-$\omega$B97X-D predictions on the reference TD-DFT scale, which lies about 0.9--1.0~eV above ADC(2) and experiment.  (a) Library design and schematic fluorescence workflow. (b) Emission energies of the test dyes, comparing ML vs TD-DFT at each method's own relaxed geometries. (c) Predicted absorption-emission relationship in the screened systems. 626 emitters with emission beyond 900 nm lie outside the plotted range. Numbered diamonds mark the three designs of panel f at their predicted positions. (d) Fraction classified as emitters for each auxochrome pair, relative to all successfully analysed structures with that pair. (e) Median emission wavelength among classified emitters for each bridge-meso combination. (f) Three selected designs and their ADC(2)/def2-SVP emission wavelengths (larger, red numbers), absorption wavelengths and oscillator strengths.}

  \label{fig:emission}
\end{figure}

Having shown that OMNI-P3x can be used for reliable screening of absorption properties out-of-the-box, we turn to the more demanding task of predicting emission properties. In addition to optimizing the ground-state structure of the system, it requires locating its minimum on the $S_1$ PES, and calculating excited-state properties there. For this purpose we selected the xanthylium cation, a molecular scaffold common for pyronin, rhodamine and rhodol dyes with numerous applications in, e.g., fluorescence labelling~\cite{Kuhn2025} and bioimaging~\cite{Farrants2024,Fiore2025}. The development of new derivatives with tailored photophysical properties remains an active area of research, driven by advances in live-cell and super-resolution imaging~\cite{reja2025,holland2026}. The chosen substitution pattern and design space it spans, shown in Figure~\ref{fig:emission}a, is based on established synthetic literature demonstrating emission-property tuning across the entire visible and near-infrared range~\cite{wang2019,lavis2017,koide2011,grimm2015,grimm2020}. Modifications include heteroatom substitution at the bridge position~(10), functional group change at the \textit{meso}~(9) position, electron donors at the auxochromic (3,6) positions, as well as donor/acceptor substitutions at the remaining ring positions~(2,4,5,7). The two halves are substituted independently, giving a total of 162,960 candidate dyes.

To refine the quality of the model predictions, we have fine-tuned the OMNI-P3x model on a small subset of the screened molecules. They were drawn at random and labelled at the reference level of theory ($\omega$B97X-D/def2-SVP/TDA), which is also the level of the excited-state data used in the pre-training of OMNI-P3x. Points were obtained by sampling frames from $S_1$ optimization trajectories, normal mode sampling around  minima, interpolations between ML and TD-DFT predicted structures, as well as single-point calculations. Full details on the fine-tuning set construction are available in the Methods section. The final set spanned about 32k conformations, and, importantly, required performing the costly $S_1$ optimization workflow for only about 1,000 molecules, less than 1\% of the screened systems.

On an independent test set of 513 dyes~(10 were excluded due to optimization ending at an $S_0/S_1$ crossing at the TD-DFT or ML level of theory), fine-tuned OMNI-P3x achieves an emission-energy MAE of 0.21 eV, while closely reproducing the reference $S_1$ minimum geometries, with a median RMSD of 0.23 \AA, as shown in Figure~\ref{fig:emission}b. Similarly to the absorption benchmarks, this comparison is made between optimized geometries by each method, measuring the discrepancies in the outcome of the full workflow. The full calculation takes an average of 12.6 s per dye on a single GPU, compared to over 5 hours with DFT on 16 CPU cores, a speed-up of about 1,500. In total, the emission screening of the full design space took 570 GPU-hours and would require about 14 million CPUh with TD-DFT. We classify the screening outcomes based on the properties of the $S_1$ state at the reached minimum: an oscillator strength $\geq$ 0.05 is labelled an emitter, and below as dark. Molecules with $S_0$/$S_1$ energy gaps below 0.5 eV were assumed to likely undergo internal conversion to the ground-state, and thus excluded from the fluorescence analysis.
A further 2,746 molecules were excluded from the analysis due to the optimization reaching an $S_0$/$S_1$ crossing point, where reference TD-DFT is not reliable~\cite{Levine2006}. The relationship between absorption and emission wavelength in the systems classified as emitters is shown in Figure~\ref{fig:emission}c. The (3,6) auxochromic positions were identified as having the strongest effect on the emissive character of $S_1$, based on fitting additive models using substitution identities as categorical inputs. The contribution of each substitution-site group was assessed from the decrease in $R^2$ when that group was omitted and the model refitted~(see Methods). Two -NH$_2$ groups give the highest predicted emitter fraction of 95\%, as shown in Figure~\ref{fig:emission}d, while the N-methylanilino~(N(Me)Ph) group strongly suppresses emission. In the class of predicted emitters, the bridge and meso substitutions are the strongest predictors of emission energy. In particular, the SO$_2$ and P(=O)Ph bridges, as well as strong electron withdrawing substituents (CF$_3$ and CN groups) give rise to the most red-shifted fluorescence, as shown in~Figure~\ref{fig:emission}e.

Additionally, we have selected 3 candidate systems with promising emission properties in the red, near-infrared~(NIR) and short-wavelength infrared~(SWIR) regimes, which are shown in panel f. To provide validation for the TD-DFT results, we re-computed the properties at ADC(2) level, using TD-DFT optimized geometries. Tested on 23 experimentally known systems,
such workflow predicts emission energies in excellent agreement with experimental emission wavelengths (MAE 0.1 eV), while TD-DFT systematically overestimates them with a mean signed error of 0.9 eV. Hence, we selected the candidates by shifting the OMNI-P3x predictions by a 0.89 eV offset, the mean ADC(2)/TD-DFT difference over the 98 emitters that completed both levels from a random sample of 100. For more details of the calibration and validation versus experiment procedure, see Section S1 of the Supporting Information. Three windows were defined on the shifted scale: red (600–700 nm), NIR (700–1000 nm), and SWIR (1000–1300 nm). Five molecules per window were selected by predicted emission oscillator strength, with at most two per bridge, and recalculated with ADC(2) single points on TD-DFT-optimized $S_1$ geometries (Table~S2). The ADC(2) emission wavelengths can fall outside the original selection windows.

\section{Conclusions}
We have presented OMNI-P3x, a universal MLIP for ground- and singlet excited-state simulations of neutral and charged organic molecules. The model combines the AIMNet2 architecture with multi-state and all-in-one learning, extending its applicability through broader elemental coverage and a newly generated dataset of ground- and excited-state conformations. The model shows increased accuracy compared to its predecessor, OMNI-P2x, both from reproduction of the reference method, as well as training data sampled using a more accurate DFT functional. It is capable of performing all the necessary steps required for typical photochemical workflows, such as simulating absorption spectra~(with or without inclusion of vibrational effects) and emission spectra in one model, carrying out the ground- and excited-state geometry optimization, frequency calculations, and prediction of optical properties with accuracy approaching TDA-$\omega$B97X-D. We benchmark this by comparing outcomes of complete workflows, supporting its capabilities as a surrogate for TD-DFT modeling. Beyond predicting energetics and optical properties, OMNI-P3x provides insight into the redistribution of electronic charge upon excitation \textit{via} learned, state-specific neural partial charges. We show that they reproduce character and direction of intramolecular charge transfer despite the absence of reference atomic charges in the training data.

We demonstrate the capabilities of the new model through two high-throughput screening workflows. First, we use the pretrained OMNI-P3x model to screen a combinatorial library of 234,375 flavylium dyes, performing ground-state geometry optimization and absorption-spectrum prediction without task-specific fine-tuning. This allows us to identify derivatives with red-shifted absorption and quantify the effects of substituent identity and position on their optical properties. We then apply fine-tuned OMNI-P3x to screening 162,960 xanthene-based candidate emitters, including relaxation to the first excited-state minimum. The screening reveals which structural modifications control the emissive character of this state and the predicted emission wavelength. Calibration and subsequent validation with ADC(2) calculations support the selection of promising candidates for red, near-infrared, and short-wavelength infrared emission.

At the same time, the accuracy of the model remains constrained by the training data. The underlying TDA-TD-DFT cannot give a reliable description of multireference systems~\cite{Levine2006} and displays systematic errors in the description of certain systems and types of excitations. Due to choosing the modestly-sized def2-SVP basis set, dictated by computational cost, Rydberg excitations remain poorly described. However, the presented all-in-one learning scheme provides a straightforward route to incorporating additional, more accurate reference levels as suitable training data become available.
\section{Methods}
\subsection{Quantum-chemical calculations}
All excited-state reference data were computed with the range-separated hybrid functional $\omega$B97X-D~\cite{chai2008} within the Tamm--Dancoff approximation (TDA)~\cite{hirata1999}, using the def2-SVP basis set~\cite{weigend2005} and the RI-J approximation~\cite{eichkorn1995,weigend2006} as implemented in TURBOMOLE~7.9~\cite{turbomole}. For every structure, we computed the energies of the closed-shell singlet ground state (S$_0$) and of the ten lowest singlet excited states, together with the transition dipole moments, and, hence, oscillator strengths, of all ten S$_0\rightarrow$S$_n$ excitations. Energy gradients were always computed for S$_0$ and S$_1$; gradients of higher excited states were added as long as they fit into a fixed budget of 10~min of computing time per structure, estimated from a cost model fitted against the number of basis functions. Dipole moments are available for S$_0$ and for every excited state whose gradient was computed. All calculations were orchestrated with MLatom~\cite{mlatom, Dral2024}.

ADC(2) results presented in figure~\ref{fig:vs_tbe} were computed with the ricc2 module\cite{haettig2005} of TURBOMOLE 7.9,\cite{turbomole} using the aug-cc-pVTZ basis set,\cite{dunning1989,kendall1992} the resolution-of-the-identity approximation with the matching auxiliary basis set,\cite{weigend2002} and the frozen-core approximation. As for all other methods in the comparison, single points were computed at the published QUEST geometries without symmetry, and the lowest excited singlet was taken as S$_1$. Twenty singlet excitations were requested for each molecule, as with fewer roots the solver could miss the lowest valence $\pi\rightarrow\pi^*$ states behind the Rydberg states introduced by the diffuse basis. The resulting S$_0\rightarrow$S$_1$ excitation energies of all 40 molecules agree with the ADC(2)/aug-cc-pVTZ values reported in QUEST~\cite{Loos2020QUEST3,Veril2021QUEST5} to within 0.001~eV.

\subsection{Data sampling}
The initial structures were taken from the database of experimentally characterized chromophores of Joung \textit{et al.}~\cite{joung2020}. After deduplication, 4,657 unique molecules were converted from SMILES to three-dimensional structures with RDKit~\cite{rdkit} and optimized with GFN2-xTB~\cite{bannwarth2019gfn2}. Ground-state conformations were sampled by GFN2-xTB metadynamics~\cite{grimme2019metadyn} (bias strength $k_\mathrm{push}=0.10~E_\mathrm{h}$, Gaussian width parameter $\alpha=0.02$) at 300~K with a 1~fs time step for 10~ps, saving a structure every 200~fs, which yielded 50 conformers per molecule. Excited-state conformations were sampled by metadynamics on the S$_1$ surface with the AM1 Hamiltonian~\cite{dewar1985} (MNDO program~\cite{mndo}), using the same bias parameters for at most 2~ps; a trajectory was stopped once the S$_1$--S$_0$ energy gap fell below 0.3~eV. The xtb program was used for ground-state metadynamics propagation, while an in-house implementation in the MLatom program was used for the $S_1$ metadynamics.

For subsequent sampling, uncertainty was quantified as the disagreement between two models: a main model trained on energies and gradients, and an auxiliary model trained on energies only~\cite{Martyka2025,hou2024}. Candidate structures were ranked by this disagreement, and the most uncertain ones were selected for labeling. The top 1\% of the ranking was discarded, to avoid sampling irrelevant, chemically implausible structures.

An initial model trained on the chromophore data was used to rank the molecules of PubChemQC~\cite{nakata2017} by uncertainty quantification~(UQ) value. The 10,000 most uncertain molecules were selected, and 20 conformers of each were generated by GFN2-xTB metadynamics. This brought the total training set size up to 515,528 structures.  Uncertainty selection utilizing a model trained on this dataset was used to select a further 100,000 PubChemQC datapoints, which were labeled at their equilibrium geometries; 73,243 of them entered the dataset. After further re-training, off-equilibrium $S_0$ and $S_1$ structures were generated using normal-mode sampling following the ANI-1ccx protocol~\cite{smith2020data}. Each molecule was optimized separately in S$_0$ and in S$_1$ with geomeTRIC~\cite{geometric}; the analytic Hessian of the respective state was computed with the model and diagonalized with PySCF~\cite{pyscf} to obtain the harmonic normal modes. For each molecule and state, five conformations were generated by displacing the atoms along all normal modes such that a total harmonic energy of $(3/2)\,N_\mathrm{at}k_\mathrm{B}T$ ($N_\mathrm{at}$ atoms, $T=300$~K) was distributed randomly over the modes, giving up to one million candidates. These were ranked by the largest per-state energy disagreement between the main and auxiliary versions of the intermediate model, and the most uncertain were labeled; 23,102 of them entered the dataset.

Further candidate geometries were drawn from the AIMNet2 training set~\cite{anstine2025}, and ranked by UQ based on the model trained on all the data collected up to this point, 500,000 most uncertain geometries were selected and labeled, with 436,530 entering the dataset. Finally, to enhance performance on small molecules,  the QM9 dataset~\cite{ramakrishnan2014} (133,885 molecules) was re-labeled at the TDA-DFT/$\omega$B97X-D/def2-SVP level of theory and added to the training data, reserving 10\% as a test set.

\subsection{Model training}
OMNI-P3x is an ensemble of four models trained with the same protocol, differing
only in the random split of the training data. The newly-sampled TDA-DFT and ANI-1ccx training sets were split at random into subtraining (90\,\%) and validation (10\,\%) sets, using a different random split for each ensemble member. Of the AIMNet2 replay set, 20,000 random structures were held out for validation~(about 1\%), with the rest used in training. All models were initialized from the published AIMNet2 $\omega$B97M-D3/def2-TZVPP parameters~\cite{anstine2025}, with the weights acting on the new state and level-of-theory inputs set to zero and the oscillator-strength head initialized randomly. Self-atomic energies were refitted for each level of theory by linear regression of the energies on the element counts, and refitting after discarding outliers outside the 2nd--98th percentiles.

The models were trained with the RAdam optimizer~\cite{liu2020radam} (weight decay $10^{-8}$) in batches of 512 entries per GPU on two nodes with four GH200 GPUs each. The training took 30--37 hours per ensemble member on the rig above. The learning rate was ramped linearly to $3\times10^{-4}$ over a five-epoch warm-up and then multiplied by 0.75 whenever the validation loss had not improved for 10 epochs, until it fell below $10^{-5}$, which took 314--401 epochs; the parameters with the lowest validation loss were retained. OMNI-P3x predictions are the averages of the energies, gradients, dipole moments and oscillator strengths of the four members.
\subsection{UV-Vis absorption spectra}
The UV/Vis absorption spectra were obtained using the standard single-point convolution approximation based on the broadening function as derived in Ref.~\citenum{Bai2020}:
\begin{equation}
    L(x)= 0.619\sum_i \frac{f_i}{\Gamma} e^{-(x-x_i)^2/\Gamma^2},
\end{equation}
where $x_i$ are the predicted transition energies, $f_i$ are their oscillator strengths, and $\Gamma$ is a pre-defined Gaussian width parameter, set to 0.3~eV.

Spectral overlap used to score the similarity between reference and ML spectra is defined as
\begin{equation}
    O = \frac{\int \min\left[L^{\mathrm{ref}}(x), L^{\mathrm{ML}}(x)\right] \mathrm{d}x}{\int \max\left[L^{\mathrm{ref}}(x), L^{\mathrm{ML}}(x)\right] \mathrm{d}x},
    \label{eq:spectral_overlap}
\end{equation}
where $L^{\mathrm{ref}}(x)$ and $L^{\mathrm{ML}}(x)$ are the reference and ML spectra from Equation~2 as functions of the photon energy $x$. $O$ is 1 for identical spectra and 0 for spectra that share no area.

NEA spectra were computed from 250 geometries per molecule, generated by Wigner sampling at 0~K following $S_0$ geometry optimization and harmonic normal-mode analysis. The workflow was performed independently with OMNI-P3x and TDA-$\omega$B97X-D/def2-SVP, using each method's own optimized minima and normal modes. Excitation energies and oscillator strengths for the ten lowest singlet excited states were evaluated at each sampled geometry and combined using the NEA implementation in MLatom~\cite{mlatom, Dral2024}, with Gaussian broadening of standard deviation $\sigma = 0.05$~eV.

\subsection{High-throughput screening of flavylium dyes}
Initial molecular structures were generated by enumerating substitutions of the flavylium scaffold with H, CN, OMe, OH and NMe$_2$. Positions 3, 5, 6, 7, 8 and $4'$ were varied independently, while the symmetry-equivalent $3'/5'$ positions were represented by the 15 unordered substituent pairs. This yielded $N = 5^6\times15=234{,}375$ distinct substitution patterns. Three-dimensional structures were generated with RDKit.

Ground-state optimizations were carried out using OMNI-P3x at the TDA-$\omega$B97X-D level descriptor. The optimizations were run using GeomeTRIC~\cite{geometric}. Excitation energies and oscillator strengths were calculated for the first four singlet excited states at each optimized geometry. Reference DFT calculations used the same settings as used in the data-generation step.

The UV/Vis absorption spectra were obtained using the same settings and approach as in the benchmark presented in Fig.~\ref{fig:spectra-UVVIS}. The first absorption maximum was defined as the lowest-energy local maximum reaching at least 2\% of the spectrum’s maximum intensity. If no qualifying local maximum was found, the global maximum was used.

The benchmark on which the error statistics in Figure~\ref{fig:absorption-screening}b were computed is based on 500 randomly selected derivatives and nine reference compounds. All errors were calculated as OMNI-P3x predictions versus reference values, using the geometries optimized independently with each method. Spectral agreement was quantified by the Spearman rank correlation between the spectra, defined as:
\begin{equation}
    r_{\mathrm{SCC}} = 1 - \frac{6}{n(n^2-1)} \sum_i d_i^2,
\end{equation}
where $d_i$ is the difference in rank between the two spectra, and $n$ is the total length of the spectrum.

For reference evaluation of screening candidates, the 100 molecules with the lowest predicted $S_0\rightarrow S_1$ excitation energies among structures containing at most four non-H substituents were selected. These candidates were evaluated with TDA-\(\omega\)B97X-D/def2-SVP at the corresponding DFT-optimized geometries.

Statistical analysis of substitution effects used the predicted vertical $S_0\rightarrow S_1$ excitation energies across the combinatorial library. The six independently varied positions were treated as categorical factors with five levels each, and the joint $3'/5'$ pair as a factor with 15 levels. Equal weight was assigned to each substitution pattern. The share of variance explained by factor $j$ was calculated as
\begin{equation} P_j=\,
\frac{\sum_{\ell}n_{j\ell}
\left(\overline{E}_{j\ell}-\overline{E}\right)^2}
{\sum_{i=1}^{N}\left(E_i-\overline{E}\right)^2} \times 100 \%,
\end{equation}
where $\overline{E}_{j\ell}$ is the mean excitation energy for level $\ell$ of factor $j$, $n_{j\ell}$ is the number of patterns with that level, and $\overline{E}$ is the overall mean. This analysis was used to describe the variance within the enumerated library, accounting for 56\%, with the remaining 44\% attributable to inter-substituent interactions. For visualization, the contribution of the joint $3'/5'$ factor was divided equally between the two positions. Substituent-induced energy shifts were calculated as the difference between the mean $S_0/S_1$ transition energy for a given substituent and the corresponding mean for H, averaging over all other substitutions elsewhere. For the pooled $3'/5'$ comparison, molecules containing the substituent at either position were compared with molecules containing H at either position. The 23 missing predictions, for which calculations failed,  were estimated using an iterative additive model for the variance decomposition; substituent averages used successful calculations only.

\subsection{High-throughput screening of xanthene-based emitters}
For fine-tuning of OMNI-P3x in the emission screening workflow, two sets of molecules were drawn from the design space, both disjoint from the test set and from each other by canonical SMILES. The purpose of the first set was to improve the ability of the model to describe $S_1$ minima of the studied dyes. It comprised 1,000 molecules, for
which the same four-step workflow as in the screen was run: $S_0$ optimization, vertical excitation (three states), $S_1$ optimization (two states) and emission at the $S_1$ minimum. The data coverage was further expanded by performing normal-mode displacements of the TD-DFT $S_1$ minima, at 50K and 200K, using OMNI-P3x frequencies. To improve particularly difficult classes of systems, for 317 molecules in which the baseline OMNI-P3x model $S_1$ minimum differed by more than 0.3~\AA{} with respect to TD-DFT, linear interpolation points were drawn between the reference and ML minima, 6 per molecule. Conformations in which the $S_1$ energy was more than 3 eV above the TD-DFT $S_1$ minimum of that  molecule were discarded from the training set.

The aim of the second set was to improve the chemical space coverage of the model. It consisted of 15k randomly drawn systems, for which cheap baseline OMNI-P3x optimizations of $S_0$ were carried out, followed by a single TD-DFT calculation of the four lowest excitations and their oscillator strengths was performed at that geometry, without energy gradients. The complete fine-tuning set contained 31765 conformations.

The fine-tuning process used a lower learning rate of $5\times10^{-5}$ after a five-epoch warm-up. The same learning rate reduction scheduler was used as in pre-training, driving the optimization process until learning rate fell below $5\times10^{-6}$. The loss weight for energies, forces, oscillator strengths and dipole moments were set to  1, 1, 0.5 and 0.1, respectively. To prevent catastrophic forgetting, a random 1\% of the original OMNI-P3x training data at each of its levels of theory was added to the training set as replay data, and the fine-tuning data were presented twice per epoch so that they made up roughly half of the training samples. A single ensemble member of OMNI-P3x was used (no. 0).

The ADC(2) calculations used the same ricc2 program settings as those used in the QUEST benchmark, and are computed at  single points at the TDA-$\omega$B97X-D/def2-SVP $S_0$ and $S_1$ minima, using the def2-SVP basis set~\cite{weigend2005}.

\subsection{Excited state charges}
TD-DFT calculations used the level of theory of the training data, as single points at the published QUEST geometries (CC3 or CCSD(T)), without symmetry. Each QUEST transition was assigned to a TD-DFT state with the same irreducible representation, determined from the orientation of the transition dipole moment, lying within $\pm0.8$ eV of the TBE/aug-cc-pVTZ value. If several states qualified, which was the case for 4 of the 26 transitions, the one with the largest dipole change was chosen. OMNI-P3x was evaluated at the same geometries with the TDA-$\omega$B97X-D level descriptor and read at the same state index as the assigned TD-DFT state.

In OMNI-P3x, the change of the atomic charges is taken directly from the model output, as the difference between the neural charges of state $S_n$ and of $S_0$ on each atom, averaged over the four ensemble members:
\begin{equation}
    \Delta q_i^{n} = q_i^{n} - q_i^{0}
\end{equation}
For TD-DFT, the relaxed difference density $\Delta\rho^{n}(\mathbf{r}) = \rho^{n}(\mathbf{r}) - \rho^{0}(\mathbf{r})$ was integrated over atomic cells:
\begin{equation}
  \Delta q_i^{n,\mathrm{ref}} = -\int_{V_i} \Delta\rho^{n}(\mathbf{r})\,\mathrm{d}\mathbf{r}
  \approx -\Delta V \sum_{k:\,\mathbf{r}_k \in V_i} \Delta\rho^{n}(\mathbf{r}_k),
  \label{eq:dq_ref}
\end{equation}
where $V_i$ is the Voronoi cell of atom $i$, i.e., all points closer to its nucleus than
to any other, $\mathbf{r}_k$ are the points of the grid, and $\Delta V$ is the volume of one grid cell, to obtain comparable per-atom charge changes.

In each of the systems, a donor~(losing electrons, more positive charge) and acceptor (gains electrons, more negative charge) fragments were identified, with a possible bridging unit connecting them. The charge transferred from donor to acceptor was calculated by summing over those fragments:
\begin{equation}
  q_{\mathrm{D}\rightarrow\mathrm{A}}^{n} = \frac{1}{2}
  \left( \sum_{i \in \mathrm{D}} \Delta q_i^{n} - \sum_{i \in \mathrm{A}} \Delta q_i^{n} \right).
  \label{eq:qct}
\end{equation}

\section{Code and data availability}
The OMNI-P3x dataset is available at \url{https://doi.org/10.6084/m9.figshare.34937758}, along with data supporting the findings: outcomes of the flavylium and xanthene screening, including the fine-tuning data and model.

 The code is available in the open-source MLatom program under the MIT license as described at \url{https://github.com/dralgroup/mlatom}, with model weights deposited on Zenodo: \url{https://doi.org/10.5281/zenodo.24344107} downloaded automatically by the program. Tutorials for running OMNI-P3x calculations in MLatom are available at \url{http://mlatom.com/docs/tutorial_omnip3x.html}.
Users can also run these calculations with natural-language instructions through our agentic system Protomia,\cite{Protomia} the successor to Aitomia.\cite{Aitomia} Protomia is available online on Aitomistic Hub\cite{AitomisticHub} and Aitomistic Lab@XMU.\cite{AitomisticLabXMU} Locally, users can run these calculations with the MLatom Skills library.\cite{mlatom-skills}

\section{Author contributions}
M.M. designed and developed the code for training the OMNI-P3x models, generated the training data, designed and performed the presented benchmarks, conceived and performed the presented screening pipelines, as well as wrote the original version of the manuscript and prepared the figures. J.J. and P.O.D. participated in the result analysis and interpretation, co-supervised the research, and revised the manuscript.

\begin{acknowledgement}

P.O.D. acknowledges funding for International Senior Scientists of the National Natural Science Foundation of China (Project No.: W2531013) and from New Generation Artificial Intelligence-National Science and Technology Major Project 2026ZD0128501.
We acknowledge Polish high-performance computing infrastructure PLGrid for awarding this project access to the LUMI supercomputer, owned by the EuroHPC Joint Undertaking, hosted by CSC (Finland) and the LUMI consortium through PLL/2026/10/019484.
This work was supported by the PIAST-AI Factory project - PIAST AIF, co-financed by the EuroHPC Joint Undertaking through the European Union's Digital Europe Programme, and by Poland as the Participating Country, under Grant Agreement No. 101250730. Views and opinions expressed are however those of the author(s) only and do not necessarily reflect those of the European Union, EuroHPC Joint Undertaking or Poland. Neither the European Union, the EuroHPC Joint Undertaking nor Poland can be held responsible for them.

M.M. and J.J. acknowledge the National Science Center, Poland, for funding this research under the program “PRELUDIUM,” grant number 2025/57/N/ST4/03587, as well as gratefully acknowledge Polish high-performance computing infrastructure PLGrid (HPC Center: ACK Cyfronet AGH) for providing computer facilities and support within computational grant no. PLG/2026/019355.

The authors acknowledge the use of Large Language Models in the preparation of this manuscript, employed for coding, numerical experiment design, and data analysis, as well as figure preparation and proofreading of the manuscript. Models used: Claude Opus 4.6-5.5 and Fable 5.0-5.2 \textit{via} the Claude Code harness, as well as GPT-6 Astra \textit{via} the Codex harness. All scientific content and final prose were thoroughly reviewed and approved by the authors.

\end{acknowledgement}

\bibliography{omni-p3x}

\begin{tocentry}
\centering
\includegraphics[width=3.23in,height=3.5cm,keepaspectratio]{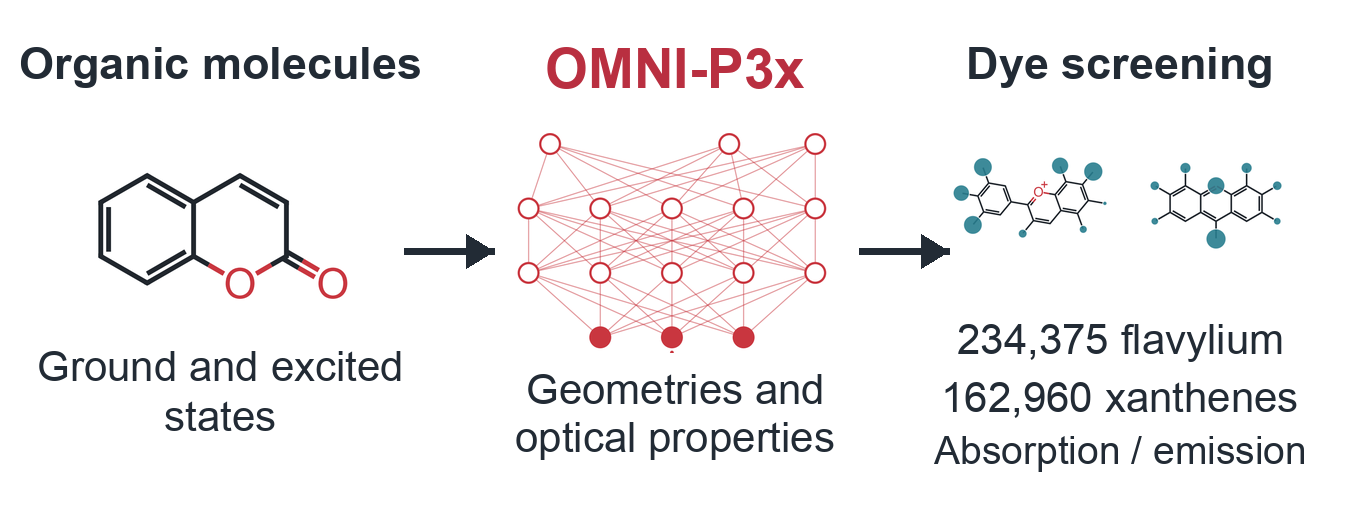}
\end{tocentry}

\end{document}